\documentclass[sigconf]{acmart}

\AtBeginDocument{%
  }

\newcommand{\ie}{\textit{i}.\textit{e}.}
\newcommand{\eg}{\textit{e}.\textit{g}.}

\usepackage{amsmath,amsfonts,bm}

\def\eqref#1{equation~\ref{#1}}

\def\1{\bm{1}}

\def\vb{{\bm{b}}}

\def\vt{{\bm{t}}}

\def\evc{{c}}
\def\evd{{d}}

\def\evn{{n}}

\def\mE{{\bm{E}}}

\def\mK{{\bm{K}}}

\def\mM{{\bm{M}}}

\def\mQ{{\bm{Q}}}

\def\mT{{\bm{T}}}

\def\mV{{\bm{V}}}
\def\mW{{\bm{W}}}

\DeclareMathAlphabet{\mathsfit}{\encodingdefault}{\sfdefault}{m}{sl}
\SetMathAlphabet{\mathsfit}{bold}{\encodingdefault}{\sfdefault}{bx}{n}

\begin{document}

\title{ProtChatGPT: Towards Understanding Proteins with Hybrid Representation and Large Language Models}

\author{Chao Wang}
\email{chao.wang-11@student.uts.edu.au}
\affiliation{%
  \institution{University of Technology Sydney}
  \city{Sydney}
  \state{NSW}
  \country{Australia}
}

\author{Hehe Fan}
\affiliation{%
  \institution{Zhejiang University}
  \city{Hangzhou}
  \state{Zhejiang}
  \country{China}
}

\author{Ruijie Quan}
\affiliation{%
  \institution{Zhejiang University}
  \city{Hangzhou}
  \state{Zhejiang}
  \country{China}
}

\author{Lina Yao}
\affiliation{%
 \institution{University of New South Wales}
 \city{Sydney}
 \state{NSW}
 \country{Australia}
}

\author{Yi Yang}
\affiliation{%
  \institution{Zhejiang University}
  \city{Hangzhou}
  \state{Zhejiang}
  \country{China}
}

\renewcommand{\shortauthors}{Wang et al.}

\begin{abstract}
Protein research is crucial in various scientific disciplines, but understanding their intricate structure-function relationships remains challenging. 
Recent advancements in Large Language Models (LLMs) have significantly improved the comprehension of task-specific knowledge, suggesting the potential for specialized ChatGPT-like systems in protein research to aid fundamental investigations. 
In this work, we introduce ProtChatGPT, which aims to learn and understand protein structures using natural language.
ProtChatGPT enables users to upload proteins, ask questions, and engage in interactive conversations to produce comprehensive answers. 
The system comprises multi-level protein encoding, protein-language alignment, and instruction tuning of LLMs.
A protein first undergoes multiple protein encoders and PLP-former to produce multi-level hybrid protein embeddings, which are then aligned through a Protein Context Gating (PCG) module with contrastive learning, and projected by an adapter to conform with the LLM.
The LLM finally combines user questions with projected protein embeddings to generate informative answers. 
Experiments show that ProtChatGPT can produce promising responses to proteins and the corresponding user questions. 
We hope that ProtChatGPT could form the basis for further exploration and application in protein research.
Code and our pre-trained model will be publicly available.
\end{abstract}





\maketitle

\section{Introduction}
Proteins, as essential molecular entities for life, hold paramount significance in biological processes. 
The comprehensive understanding of protein structure and function is important for advancing research in the realms of biology and biomedicine.
However, traditional protein research normally involves labor-intensive laboratory experiments and extensive literature reviews, which could be time-consuming and require specialized expertise in protein.

Recently, Large Language Models (LLMs), \eg, ChatGPT~\citep{GPT}, have prevailed in Natural Language Processing (NLP)~\citep{T5,palm,llama}.
With superior language understanding and logical reasoning capabilities, these models can perform various intricate linguistic tasks such as question and answering (Q\&A)~\citep{miniGPT4,BLIP2}. 
Considering that primary sequences can be regarded as protein's own ``natural language'', this intuitively motivates us to ride on LLMs' coattails and customize them into protein research based on large-scale biological corpora, \eg, RCSB-PDB~\citep{ProteinChat}.

Empirically, with the capabilities of LLMs specialized in protein, researchers can potentially achieve \textit{(1) Protein Understanding and Analysis} by simplifying the retrieval of crucial information (\eg, structures, functions, interactions, mutations, and disease associations) about specific proteins for research; \textit{(2) Customized Protein Design} by characterizing the unique protein structures of patients to discover targeted drugs and further verify expected functions for healthcare.

\begin{figure*}[t]
    \centering
    \includegraphics[width=0.98\textwidth]{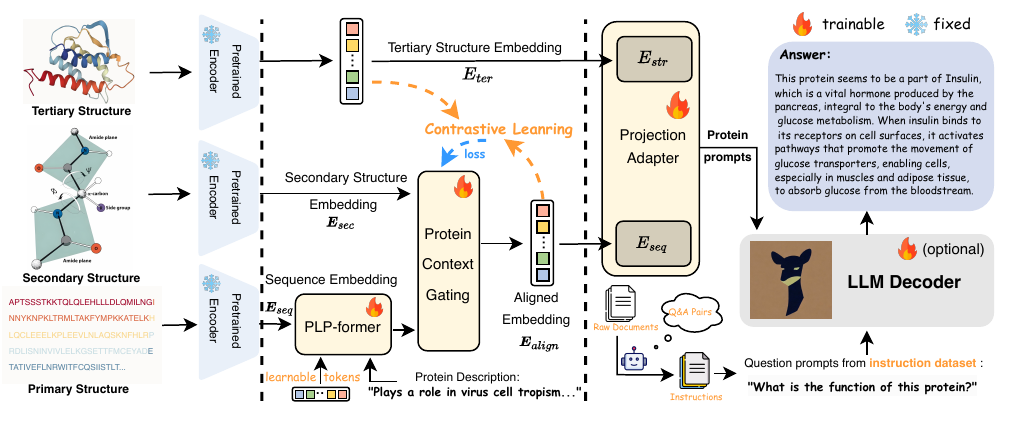}
    \vspace{-4mm}
    \caption{Overview of the ProtChatGPT framework. Our pipeline consists of three stages: (1) multi-level protein encoding, (2) multi-level protein-language alignment, and (3) instruction tuning with an external web corpus and protein features.
    First, we utilize three pre-trained frozen large protein encoders to acquire high-quality multi-level embeddings.
    In the second stage, we first enforce the PLP Transformer, a lightweight transformer with learnable query tokens, to learn the protein representation most relevant to the text description.
    The PLP-former takes the sequence embedding $\mE_{seq}$, tokens, and protein descriptions as inputs and outputs the learned tokens as the selected embedding.
    The selected embedding is first aligned with the secondary structure embedding $\mE_{sec}$ through protein context gating for aligned embedding $\mE_{align}$, and then we adopt contrastive learning to enforce the tertiary structure embedding $\mE_{ter}$ to further align with the joint representation $\mE_{align}$.
    In the third stage, we perform protein-to-text generative learning by connecting the aligned $\mE_{align}$ and $\mE_{str}$ to an LLM decoder.
    Combined with the instruction pairs extracted from open protein-related literature, an adapter is further trained as the information bottleneck between protein embeddings and the LLM, such that its output can be interpreted by the language model. 
    Finally, the LLM can produce descriptive answers given the question prompt and the multi-level protein prompt from the adapter.
    }
    \label{fig:teaser}
    \vspace{-2mm}
\end{figure*}

In this paper, we propose an AI-based protein chat system, named \textbf{ProtChatGPT}, to implement ChatGPT-like functionalities for the protein research field.  
ProtChatGPT works in a similar way to natural language conversation systems.
Users are allowed to upload protein 1D sequences or 3D structures (\eg, fasta or pdb files) and pose diverse related questions. 
Then, ProtChatGPT produces comprehensive responses in an interactive manner based on the questions. 
In this way, researchers can intuitively acquire valuable insights and interactively delve into the complexities of diverse proteins. 
Specifically, ProtChatGPT consists of three stages: multi-level protein encoding, protein-language alignment, and instruction tuning of LLMs, as shown in Figure~\ref{fig:teaser}. 
First, we employ three pre-trained protein encoders to embed the primary sequence (\ie, by ESM-1b~\citep{ESM-1b}), secondary structure (\ie, by NetSurfP~\citep{NetSurfP}), and tertiary structure (\ie, by ESM-IF1~\citep{ESM-IF}), respectively. 
Next, to align the protein sequence and language modalities, we propose the PLP-former. 
PLP-former extracts features from the output of the protein encoder, and learns the protein representations that are most relevant to the text description. 
We also deploy a Protein Context Gating (PCG) module trained with contrastive loss further to align the sequence with other levels of protein embeddings.
Third, we use an adapter as an information transmitter to convert protein embeddings into protein prompts that can be interpreted by the LLM.
Finally, the LLM combines user questions (\ie, question prompts) with the transmitted protein prompts to produce corresponding answers. 
We conduct experiments on protein understanding and design. 
Experimental results show the effectiveness of the proposed method. 
In summary, our contributions are as follows:
\begin{itemize}

    \item We propose ProtChatGPT, an interactive ChatGPT-like system that engages Q\&A for protein-related research, which significantly facilitates protein understanding and design.
    \item We introduce a progressive protein-language learning strategy for dynamically aligning protein from multi-level structures with natural language, including a transformer-based PLP-former and a Protein Context Gating (PCG) module based on contrastive learning.
    \item As a minor contribution, a protein-specific instruction tuning dataset is built to facilitate the LLMs to better correspond to the user questions.
    \item We demonstrate ProtChatGPT’s versatility and range of applications by deploying it to tasks of a rather distinct nature, including protein understanding and design.
    
\end{itemize}

\section{Related Work}
\textbf{Protein Representation Learning.}
Proteins are workhorses of the cell, which contain four distinct levels of structures carrying their fundamental functions.
Previous protein representation works seek to learn protein representations based on different levels of structures. 
Considering protein sequences as language, several works~\citep{madani2023large,notin2022tranception} encode amino acid tokens using transformer model \citep{transformer} to extract pairwise relationships among amino acids, and autoregressively recover protein sequences based on extensive protein sequence databases.
Alternatively, other sequence modeling methods\citep{ESM-2,ESM-1v,ESM-1b,ESM,vig2020bertology} resort to using Masked Language Modeling (MLM) to develop attention patterns that correspond to the residue-residue contact map of the protein.
Compared with sequence-based methods, structure-based methods~\citep{gligorijevic2021structure,GVP,GearNet} directly encode geometric information of proteins for topology-sensitive tasks such as molecule binding~\citep{jin2021iterative, kong2022conditional}, protein interface analysis \citep{mahbub2022egret, reau2023deeprank}, and protein properties prediction \citep{zhang2022ontoprotein}. 
In this paper, we aim to directly leverage these pre-trained Large Protein Models (LPMs) to acquire high-quality embeddings without fine-tuning their network parameters.

\begin{figure*}[t]
	\centering
	\includegraphics[width=0.7\textwidth]{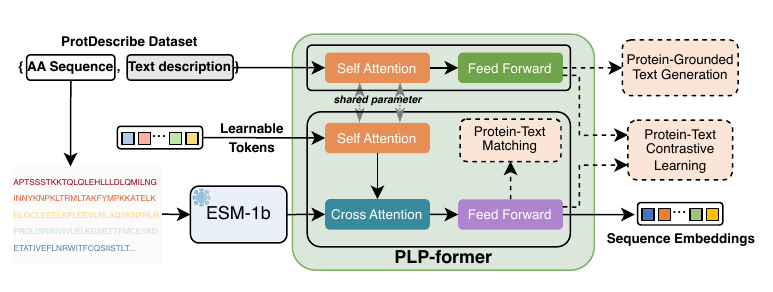}
	\vspace{-4mm}
	\caption{Illustration of the PLP-former and protein-language representation learning. PLP-former consists of two transformer submodules with shared self-attention: (1) a text transformer that performs encoding and decoding of protein descriptions, and (2) a protein transformer that interacts with the frozen ESM-1b for sequence feature extraction. PLP-former is trained by jointly optimizing three pretraining objectives (dashed boxes) on sequence-description pairs.}
        \label{fig:PLP-former}
	\vspace{-2mm}
\end{figure*}

\textbf{Large Language Models (LLMs).} 
Recently, Natural Language Processing (NLP) has witnessed significant advances due to the development of LLMs~\citep{GPT,bert} trained on an extensive, diverse dataset. 
Consequently, many multi-modal variants~\citep{galactica,Flamingo,GPT-4,ChatCAD,MedVQA} based on LLMs have gained significant attention for the understanding of information in other modalities beyond text.
For example, BLIP-2\citep{BLIP2} designs a Q-Former to align the visual features from the frozen visual encoder with large language models. 
FROMAGe~\citep{FROMAGe} freezes the LLM and visual encoders, and fine-tunes linear mapping layers to achieve cross-modality interactions.
For protein understanding, MedVQA~\citep{MedVQA} employs a Multi-Layer Perceptron (MLP) network that maps the extracted visual features from a frozen vision encoder to a set of learnable tokens, which develops an open-ended VQA for diagnoses and treatment decisions.
Galactica~\citep{galactica} explicitly models the protein sequences and SMILES with scientific literature, and enables the model to explain the properties of the sequences. 
ProteinChat~\citep{ProteinChat} further models the protein structure with its corresponding descriptions in Protein Data Bank (PDB) for the protein Q\&A task. However, this method only considers the impact of protein structure on its function, while neglecting the complementarity between different levels of structure, such as amino acid sequences (primary structure) and geometric coordinates (tertiary structure).
In this paper, we aim to adapt pre-trained general LLMs (\eg, Vicuna~\citep{vicuna} and LL{\scalebox{0.75}[0.75]{A}}MA3) for protein-specific ChatGPT-like tasks, which aligns multi-level protein features from LPMs with LLMs.

\section{Proposed Method}

While demonstrating excellent performance in natural language tasks, LLM cannot directly facilitate protein question-answering tasks due to the modality gap between protein structures and biomedical descriptions.
As shown in Figure~\ref{fig:teaser}, in order to bridge this gap, we introduce a progressive protein-language learning framework with three stages:
(1) multi-level protein encoding, (2) multi-level protein-language alignment, and (3) instruction tuning with protein features.

\subsection{Multi-Level Protein Encoding}
In this paper, we leverage three pre-trained Large Protein Models (LPMs) exclusively trained on different levels of protein structures, working synergistically to provide high-quality hybrid embeddings without fine-tuning their network parameters.

First, we use a pre-trained sequence encoder to extract protein sequence features. 
Specifically, given a protein sequence with $\evn$ amino acids, the encoder produces the corresponding \textbf{sequence embedding} $\mE_{seq} \in \mathbb{R}^{\evn \times \evc_{seq}}$, where $\evc_{seq}$ is the number of embedding channels. 
In our implementation, we use ESM-1b~\citep{ESM-1b} as the sequence encoder, where $\evc_{seq}$ = 768.
Although ESM-1b is able to implicitly capture structural contact information, incorporating detailed structure (secondary and tertiary) information explicitly can be an effective way to model spatial interactions between residues. 
Therefore, we propose to enhance the ESM-1b with a supplementary protein secondary structure encoder NetSurfP~\citep{NetSurfP}.
Specifically, this encoder consists of two parallel convolution layers and an identity layer, whose outputs are concatenated in the feature dimension and fed to a two-layer bidirectional LSTM containing 1024 hidden units. The output is then projected to an 8-dimensional feature vector at each position as the \textbf{secondary structure embedding} $\mE_{sec} \in \mathbb{R}^{\evn \times 8}$. 
Moreover, we further introduce the ESM-IF1~\citep{ESM-IF} encoder for protein tertiary structure representation. 
Specifically, we select the feature from an intermediate layer as the \textbf{tertiary structure embedding} $\mE_{ter} \in \mathbb{R}^{\evn \times \evc_{ter}}$ where the number of embedding channels $\evc_{ter} = 512$, indicating the geometric protein knowledge.

By acquiring such multi-level hybrid property information, the quality of protein embeddings can be further improved, considering that the protein properties studied in ESM-1b and ESM-IF1 can correlate with each other.
Note that all the employed encoders are frozen for efficient training.

\subsection{Multi-Level Protein-Language Alignment}

\subsubsection{Protein-Language Pretraining (PLP)}\label{PLPsec}
As mentioned before, the key challenge of transferring LLMs to protein research lies in the modality gap between protein structures and biomedical texts.
Despite the strong language generation and zero-shot transfer abilities of LLMs, directly retraining them end-to-end for protein specialization appears to be impractical due to the massive number of parameters and data requirements.
Another alternative is fine-tuning the pre-trained parameters, but this often leads to catastrophic forgetting.
Considering this trade-off, we propose a Protein-Language Pretraining Transformer (PLP-former) for efficient cross-modal alignment between protein and text, while remaining LLMs frozen during the training.
Following existing Vision-Language Pretraining (VLP) works~\citep{BLIP2,miniGPT4,InstructBLIP}, we introduce the PLP-former to extract protein-related features from a frozen protein sequence encoder.
More related works on VLP are also given in Appendix~\ref{VLP}.

As shown in Figure~\ref{fig:PLP-former}, PLP-former consists of three inputs: initial sequence embedding $\mE_{seq} \in \mathbb{R}^{\evn \times 768}$ from ESM-1b, a set of learnable tokens $\mT \in \mathbb{R}^{32 \times 768}$, and the corresponding description.
The learnable tokens $\mT$ first perform mutual interactions via self-attention layers in the protein transformer.
Specifically, the tokens $\mT$ first acquire queries $\mQ$, keys $\mK$ and values $\mV$ through three linear transformation matrices $\mW_q$, $\mW_k$ and $\mW_v$. It can be formulated as: 
\begin{align}
\mQ = \mT \mW_q,
\mK = \mT \mW_k,
\mV = \mT \mW_v.
\end{align}
Next, the attention map $\mM$ is computed through $\mathrm{softmax}(\frac{\mQ\mK^T}{\sqrt{\evd_k}})$ where $\evd_k$ represents the dimensionality of the keys, \textit{softmax} is the softmax activation function.
The refined tokens $\mT' \in \mathbb{R}^{32 \times 768}$ after self-attention can be calculated as $\mT' = \mW_o \mM \mV$ where $\mW_o$ is the projection matrix for output.
Given the sequence features from ESM-1b as $\mE_{seq} \in \mathbb{R}^{\evn \times 768}$, tokens $\mT' \in \mathbb{R}^{32 \times 768}$ then interact with  language embedding through cross-attention layers, which can be formulated as:
\begin{equation}
    \mT'' = \mW_o' \left(~\mathrm{softmax}\left(\frac{\mT'\mW_q' (\mE_{seq}\mW_k')^T}{\sqrt{d_k}}\right)\left(\mE_{seq}\mW_v'\right)\right),
\end{equation}
where $\mT'' \in \mathbb{R}^{32 \times 768}$ represents the refined tokens after cross-attention, $\mW_q'$, $\mW_k'$, $\mW_v'$ and $\mW_o'$ are a new set of learning transformation matrices.
Additionally, benefiting from the text transformer, tokens can further interact with the textual descriptions through the same self-attention layers.
Finally, the PLP-former produces the output $\mE_{seq}$ after a linear feed-forward layer~\citep{transformer}.
The size of selected $\mE_{seq}$ ($32 \times 768$) is much smaller than the size of frozen sequence features ($\evn \times 768$) as the length of protein $\evn$ usually numbers over hundreds.
Depending on the pretraining task, we implement distinct attention masking strategies~\citep{BLIP2} within the self-attention block to regulate the token-text interaction. 
The training details of PLP-former are given in Appendix~\ref{PLPdetail}.
In this way, PLP-former can effectively select the most useful information for the LLM while removing irrelevant protein information. 
This reduces the burden of the LLM to learn protein-language alignment while mitigating the catastrophic forgetting problem, as the PLP-former establishes a fundamental correlation between the primary structure and its descriptive text before introducing more complex structural data.

\subsubsection{Multi-Level Protein Alignment}\label{MPAsec}
After multi-level encoding, the next step involves aligning proteins with these multi-level features to harness the full potential of their hybrid structural representations. 
In this work, we introduce a progressive alignment strategy across different levels.  

\paragraph{Protein Context Gating (PCG)}
Initially, we utilize a Protein Context Gating (PCG) module for the feature aggregation between the sequence embedding $\mE_{seq} \in \mathbb{R}^{32 \times 768}$ and secondary structure embedding $\mE_{sec} \in \mathbb{R}^{\evn \times 8}$. 
Our goal is to transform $\mE_{sec}$ into dynamic gating maps, which effectively allows the PCG module to highlight or suppress $\mE_{seq}$ based on the relevant secondary structural information.
The formulation of PCG can be written as:
\begin{equation}
    \mE_{align} = \sigma(\mW_{sec} \mE_{sec} + \vb) \odot \mE_{seq},
\end{equation}
where $\odot$ represents the element-wise multiplication, $\sigma$ denotes the sigmoid activation function that normalizes the gating values to the range [0,1]. 
A convolution layer is first deployed on $\mE_{sec} \in \mathbb{R}^{n \times 8}$ to project the dimension to $32 \times 8$.
$\mW_{sec} \in \mathbb{R}^{8 \times 768}$ is a transformation matrix, $\vb \in \mathbb{R}^{1 \times 768}$ is the bias to acquire the aligned embedding $\mE_{align} \in \mathbb{R}^{32 \times 768}$.

\paragraph{Contrastive Learning}
Subsequently, we employ a contrastive loss to achieve alignment with the tertiary structure $\mE_{ter}$, which can be formulated as:
\begin{equation}
    \mathcal{L} = -\log \frac{\exp(\mE_{align} \cdot \mE_{ter_{+}} / \tau)}{\sum_{i=0}^{k} \exp(\mE_{align} \cdot \mE_{ter_{i}} / \tau)},
\end{equation}
where $\mE_{ter_{+}}$ is the positive sample corresponding to $E_{align}$, $\cdot$ represents the dot product, $\tau$ represents a temperature hyper-parameter modulating the learning sensitivity.
We also added a convolutional layer on $\mE_{ter} \in \mathbb{R}^{n \times 512}$ to match the dimensions with $\mE_{align} \in \mathbb{R}^{32 \times 768}$.
By introducing the tertiary structural information, this contrastive mechanism allows for dynamic protein feature alignment and optimizes the representation for tasks requiring high discernment between similar protein sequences, such as homologous protein analysis.

\subsection{Instruction Tuning with Protein Features}\label{ITsec}
\subsubsection{Projection Adapter}
During the final instruction tuning stage, we further design a multi-level projection adapter to harvest the generative language capability of LLMs for protein-to-text generative learning.
The adapter takes the pre-aligned embeddings $\mE_{align}$ and $\mE_{ter}$ from ESM-IF1 as inputs, and acts as an information bottleneck to the LLM decoder, such that its output protein representation can be interpreted by the LLM.
In practice, we use two individual Fully-Connected (FC) layers to linearly project the output protein embeddings into the same dimension as the question embedding of the LLM. 
They function as soft protein prompts that condition the LLM on protein representation from different levels. 

\subsubsection{Protein-Text Generation}
Finally, the projected protein prompts are prepended to the question prompts (text embeddings of user questions) through concatenation. 
In implementation, we separately deploy two LLMs Vicuna-13B~\citep{vicuna} and LL{\scalebox{0.75}[0.75]{A}}MA3. We employ the LLM decoder to model the conditional generation probability $p_{\theta}(\vt_i|\vt_{<i})$ in the language model, where $\vt_i$ is the generated tokens, $\vt_{<i}$ represents the context vector of input tokens (\ie, protein embeddings along with user questions).
To enhance the model training with protein-text pairs, we utilize a specialized token prompt as:
\[
\centering
\colorbox{gray!15}{
$\begin{aligned}
\mathbf{Q}{:}&<{\text{Protein}}><\text{ProteinPrompts}></\text{Protein}> \\
&<\text{QuestionPrompts}> \\
\mathbf{A}{:}&<\text{Descriptions}>
\end{aligned}$
}
\]
Here $<\text{ProteinPrompts}>$ represents the soft prompts that symbolize the multi-level embeddings aligned after the projection adapter. 
$<\text{Protein}>$ and $</\text{Protein}>$ respectively represent the start and end symbols for protein embeddings.
$<\text{QuestionPrompts}>$ represents the user questions that prompt the LLM to generate corresponding answers for the uploaded protein, such as ``Describe the function of this protein''.
To enable LLM to better understand user instructions and enhance the effectiveness and accuracy of protein analysis, we created a protein instruction tuning dataset in which each protein is associated with ten Q\&A pairs (see Section~\ref{dataset} for more details).
$<\text{Descriptions}>$ represents the generated answers, which have been substituted with publicly available protein descriptions during the second training stage. 
During the tuning stage, the generated tokens ($<\text{Descriptions}>$) are replaced with the descriptions from our instruction dataset. 
During the testing and inference stage, the generated tokens remain empty, expecting ProtChatGPT to generate descriptive answers for the given protein and corresponding questions.

\section{Experiments}
\subsection{Dataset}\label{secdataset}
\subsubsection{Protein-Language Pretraining Dataset}
In order to train our ProtChatGPT, dedicated protein-specific training datasets are indispensable for our proposed progressive training strategy.
First, we adopt ProtDescribe dataset~\citep{protst} to train the PLP-Transformer for protein-description representation learning, which contains 553,052 aligned pairs of protein sequences and textual descriptions such as protein names, functions, families, subcellular locations, etc.

\subsubsection{Instruction Tuning Datasets}\label{dataset}
Nevertheless, despite covering the protein sequence positions, ProtDescribe de facto simply relies on textual descriptions to provide a rough indication of protein structural and functional similarity. 
It might be more straightforward to directly utilize structural information, especially considering that NetSurfP and ESM-IF are specifically designed for high-level protein structures.
Considering this problem, we resort to the RCSB-PDB dataset~\citep{ProteinChat} which comprises 143,508 tertiary structure-description aligned pairs of proteins. 
We further extract the protein sequence and expand this dataset with sequence and secondary structure labels.
All secondary structure labels are constructed using DSSP (Dictionary of Protein Secondary Structure, see Appendix~\ref{Proteins}).

Moreover, considering fine-tuning LLMs on other domains typically requires many task-specific examples, we further collect the corresponding literature from the RCSB PDB website based on the protein entry ID, and then extract the PubMed abstracts as an additional \textbf{external web corpus}.
Given the protein-related document, we utilize a recent method Bonito~\citep{Bonito} to convert unannotated text into corresponding Q\&A-specific instruction pairs.
In summary, for every protein taken into account, we compile its residue sequences, secondary structure label, tertiary atomic coordinates, along with corresponding 10 paired Q\&A descriptions.
We randomly select 1,000 pairs of protein for evaluation, and the rest part of the dataset is used for multi-level protein alignment (Section~\ref{MPAsec}) and instruction tuning on LLMs (Section~\ref{ITsec}).
More details and examples of our training datasets are given in Appendix~\ref{appedix:dataset}.

\subsection{Implementation Details}\label{impl}
In this paper, we use a progressive strategy to ensure that the pre-trained models retain their learned knowledge while fine-tuning them to align the protein embeddings with the LLMs.
Firstly, we freeze all the multi-level protein encoders, solely focusing on training the PLP-former with the ProtDescribe dataset.
Then the protein context gating module is trained with the RCSB-PDB dataset through contrastive learning.
Finally, we freeze the LLM decoder and train the projection adapter with the generated instruction dataset.
More implementation details are given in Appendix~\ref{implementation}.

\subsection{Qualitative Comparisons}

\begin{figure*}[t]
    \centering
    \includegraphics[width=\textwidth]{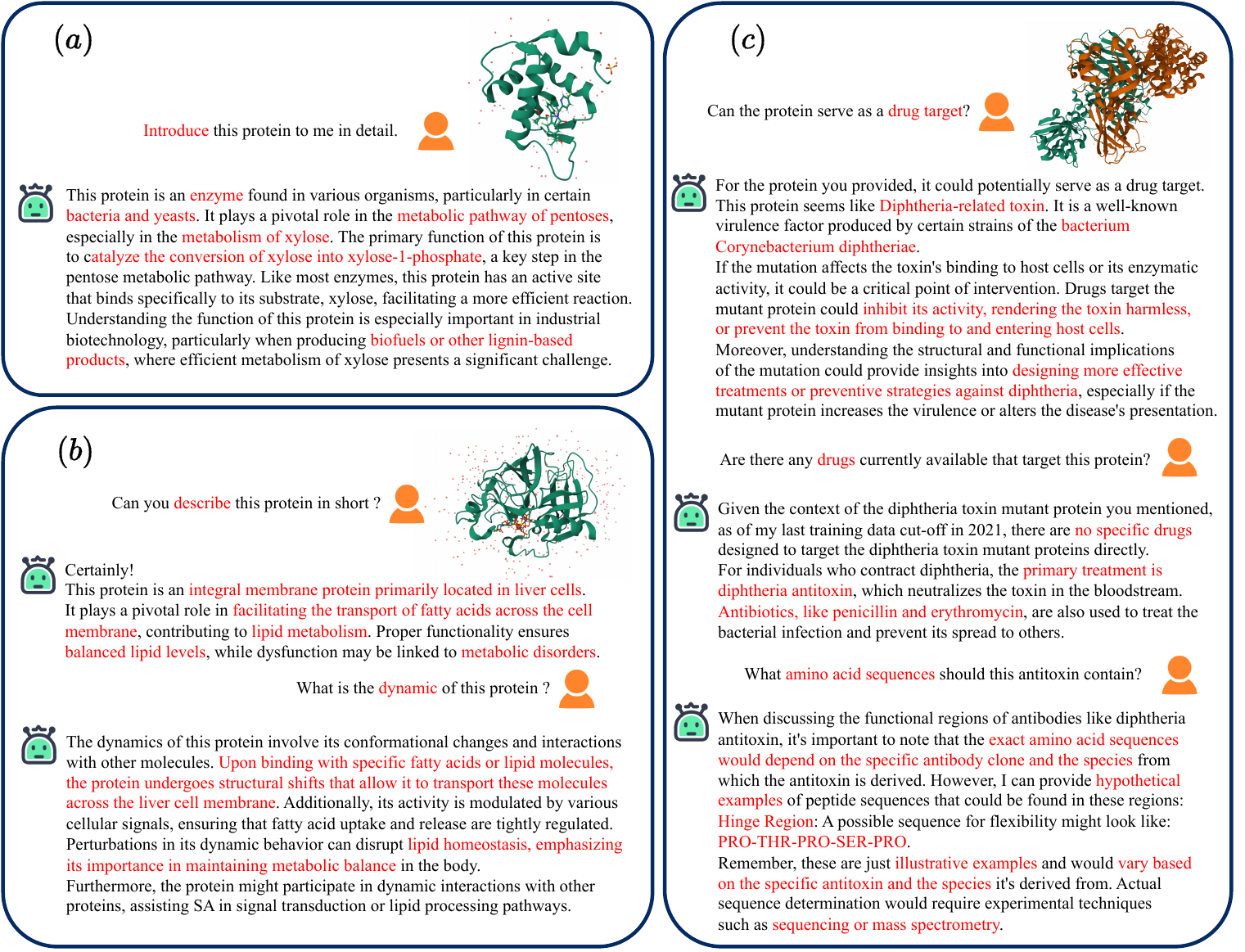}
    \caption{
    Several conversation examples. (a, b): ProtChatGPT can fully explore the intrinsic properties of proteins and accurately understand user queries, enabling protein understanding and analysis.
    (c): ProtChatGPT also has the potential to assist in drug development through pathogenicity analysis, diagnostic simulation, and protein design.
    }
    \label{fig:conversation}
\end{figure*}

We first show some example conversations between the user and ProtChatGPT to indicate the promising results on both protein understanding and design tasks.
For a fair comparison, we use the protein that does not appear in the training set of ProtChatGPT.
Figure~\ref{fig:conversation} shows several conversation examples of our ProtChatGPT related to protein understanding and design, as well as customized protein design tasks.
It can be observed that our system comprehends the meaning of the question well, providing accurate and logically consistent responses that not only identify the types and characteristics of proteins, but also extend to exploring their related application areas and potential values. 
Additionally, when presented with a protein, ProtChatGPT can provide explanations regarding questions associated with protein understanding. 
This implies that researchers can swiftly survey the structures, functions, and other relevant properties through ProtChatGPT. 
Coupled with this comprehension capability, further judgments on mutations and disease association can be made about specific proteins, leading to the potential of targeted protein design in healthcare.

\begin{figure*}[t]
	\centering
	\includegraphics[width=\textwidth]{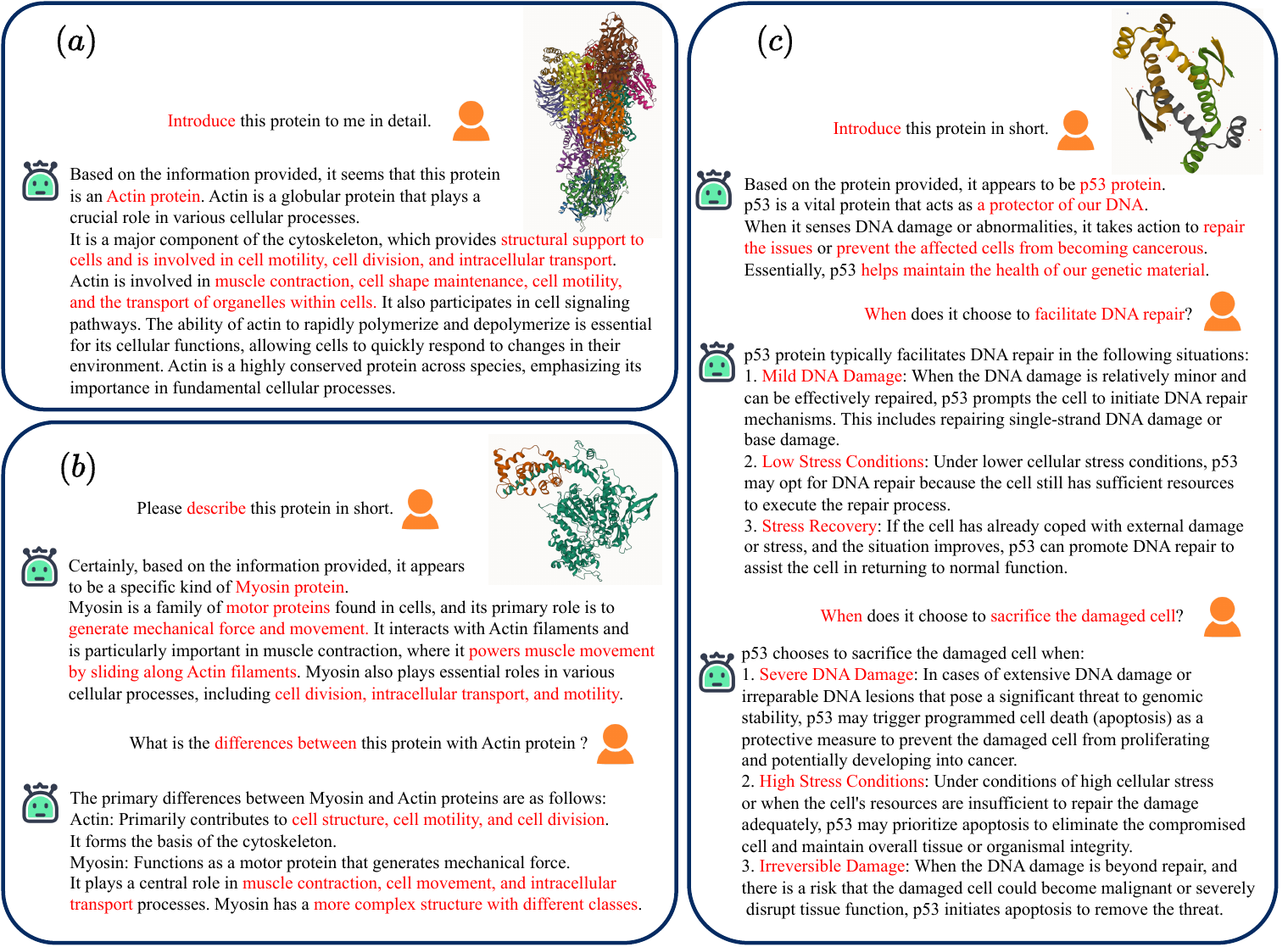}
	\caption{Challenging case studies. Benefitting from contrastive-based feature alignment and domain-specific instruction tuning, our proposed ProtChatGPT is capable of handling complex scenarios such as (a, b) homologous proteins and (c) mutually exclusive functions.}
        \label{fig:case}
\end{figure*}

\subsection{Case Study}\label{case}
To further validate the utility of our method, we conduct case studies on two specific challenging examples, including homologous proteins and mutually exclusive functions.

\subsubsection{Homologous Proteins}
Homologous proteins are proteins that are derived from the same ancestral gene, which usually share similar amino acid sequences and structures. 
This study could help in understanding evolutionary relationships, predicting protein functions, and identifying potential therapeutic interventions.
In this case, we choose \textbf{\textit{Actin}} and \textbf{\textit{Myosin}} proteins for our test. 
As shown in Figure~\ref{fig:case} $(a)$ and $(b)$, although these two proteins have similar amino acid sequences, our multi-level approach can still differentiate and make reasonable analyses due to their significant structural differences.

\subsubsection{Mutually Exclusive Functions}
Mutually exclusive protein functions describe situations where a certain protein in the same cell cannot simultaneously perform its roles, especially in signaling pathways, regulatory mechanisms, or cellular processes, where one activity inhibits or prevents another.
Understanding these functions is important for grasping how cells make decisions and respond to their environment, particularly in complex processes like development, immune responses, and disease progression.
In this case, we choose the \textbf{\textit{p53}} protein for our test. 
Figure~\ref{fig:case} (c) indicates that our method can efficiently incorporate contextual semantics to provide varied analyses under different environment prompts.

\subsection{Quantitative Comparisons}\label{quan}

\subsubsection{Protein Q\&A}
First, we provide comparisons for the protein Q\&A task on the PDB-QA dataset~\citep{ProteinChat}. We compared our LL{\scalebox{0.75}[0.75]{A}}MA3-8B based method with ProteinChat~\citep{ProteinChat} (Vicuna-13B based) and ProtT3~\citep{ProtT3} (Galactica-1.3B based). 
We followed the same test setting as ProtT3 and selected the exact match as the evaluation metric. SP stands for structure/property, SI stands for supplementary information. 
We directly used the public checkpoints. 
As shown in Table~\ref{QA}, our method surpasses other baselines due to the combined encoding of all three levels of protein structures. 
In contrast, ProteinChat only uses tertiary structure while ProtT3 only uses protein sequence information, which ignores the cross-level complementarity and may lead to insufficient protein feature extraction.

\begin{table}[h]
    \centering
    \caption{Protein Q\&A task on PDB-QA dataset with \textit{GT} and \textit{pseudo} structures. Best performances are marked in \textbf{bold}.}
    \renewcommand\tabcolsep{4pt}
    \vspace{-2mm}
    \begin{tabular}{lcccccc}
        \toprule
        \textbf{Model} & \multicolumn{2}{c}{\textbf{String}} & \multicolumn{2}{c}{\textbf{Number}} & \textbf{Overall} \\
        & SP & SI & SP & SI &  \\
        \midrule
        ProteinChat & 7.2 & 10.3 & 28.9 & 19.2 & 15.5 \\
        ProteinChat (pseudo label) & 7.8 & 9.5 & 27.4 & 17.5 & 14.2 \\
        ProtT3 & 85.4 & 69.8 & 47.2 & 39.4 & 65.0 \\
        \textbf{Ours} & \textbf{88.1} & \textbf{73.3} & \textbf{49.6} & \textbf{42.7} & \textbf{68.8} \\
        \textbf{Ours} (pseudo label) & 87.6 & 72.8 & 49.2 & 41.9 & 67.7 \\
        \bottomrule
    \end{tabular}
    \label{QA}
    \vspace{-3mm}
\end{table}

Considering that users might only have the amino acid sequence, pseudo secondary and tertiary structures can be generated using structural prediction models such as AlphaFold2~\citep{AlphaFold}.
To simulate real-world application scenarios, we further provide more experiments on a large evaluation benchmark for the Protein Q\&A task to verify the robustness under insufficient input. 
The original experiment was conducted on the PDB-QA dataset, in which the ground truth tertiary structures are directly obtained from RCSB-PDB~\citep{ProteinChat}. 
With the same test set, here we use the protein sequence to predict the pseudo tertiary structure through AlphaFold2. 
We also use the pseudo-secondary structures obtained from NetSurfP~\citep{NetSurfP} for experiments.
It can be observed that even when using pseudo-structure labels as test inputs, our model still performs well. 
Additionally, we also conducted validation experiments on previous methods (ProtT3 is not applicable since it only utilizes sequence information) using the same pseudo inputs, which also resulted in good performance. 
This further demonstrates the feasibility of using the AlphaFold dataset~\citep{varadi2022alphafold} as pseudo labels for inference.

\subsubsection{Protein-text Retrieval}
We also provide experiments on the cross-modal protein-text retrieval task, as shown in Table~\ref{re}.
We use the ProteinKG25~\citep{Ontoprotein} testset and select ProtST~\citep{protst}, ProteinCLAP~\citep{liu2023text} and ProtT3~\citep{ProtT3} for comparisons. 
As the ProteinKG25 dataset only contains amino acid sequences as the protein query, this retrieval task essentially tests the performance of our PLP-former.
For a fair comparison, we follow the same retrieval setting as ProtT3, which uses PTC to obtain the top-k ranked candidates and PTM for re-ranking (see Appendix~\ref{PLPdetail} for PTC and PTM).

\begin{table}[h]
    \centering
    \vspace{-2mm}
    \caption{Protein-text retrieval task on ProteinKG25 dataset.}
    \renewcommand\tabcolsep{4pt}
    \vspace{-2mm}
    \begin{tabular}{lcccc}
        \toprule
        \textbf{Model} & \multicolumn{2}{c}{\textbf{Protein to Text}} & \multicolumn{2}{c}{\textbf{Text to Protein}} \\
        & \textbf{Acc} & \textbf{R@20} & \textbf{Acc} & \textbf{R@20} \\
        \midrule
        ProtST & 5.5 & 41.6 & 5.8 & 43.4 \\
        ProteinCLAP & 39.0 & 89.4 & 39.3 & 89.7 \\
        ProtT3 & 55.8 & 91.7 & 55.6 & 91.7 \\
        Ours & 60.4 & 93.3 & 60.1 & 93.3 \\
        \bottomrule
    \end{tabular}
    \vspace{-2mm}
    \label{re}
\end{table}

Note that different from ProtT3 which only uses protein sequences as input, our ProtChatGPT introduces multi-level alignment from primary to tertiary structures, which leads to better protein representation (see Table~\ref{QA}), and further enables protein cross-level structure retrieval through subsequent contrastive learning (see Appendix~\ref{moreex}).

\renewcommand\arraystretch{1.2}
\begin{table*}[t] 
\caption{Quantitative comparisons on the proposed multi-level protein encoding, PCG module and PLP-former with frozen LLMs. $\uparrow$ indicates that a higher value corresponds to better performance.
	}\label{tab:ex1}
	\renewcommand\tabcolsep{6pt}
	\centering
 \vspace{-2mm}
\begin{tabular}{l|ccccccc}
\hline
\textbf{Method}     & \textbf{BLEU-1 $\uparrow$} & \textbf{BLEU-4 $\uparrow$} & \textbf{ROUGE-L $\uparrow$} & \textbf{METEOR $\uparrow$} & \textbf{CIDEr $\uparrow$} & \textbf{SPICE $\uparrow$} & \begin{tabular}[c]{@{}c@{}}\textbf{PubMed}\\ \textbf{BERTScore $\uparrow$}\end{tabular} \\ \hline
\textit{w/o} tertiary structure  & 0.461  & 0.316  & 0.411   & 0.247  & 0.515 & 0.240 & 0.348            \\
\textit{w/o} secondary structure \& PCG  & 0.509  & 0.332  & 0.445   & 0.250  & 0.544 & 0.265 & 0.411            \\
\textit{w/o} PLP-former & 0.588  & 0.356  & 0.469   & 0.278  & 0.582 & 0.287 & 0.431            \\
\textit{w/o} instruction tuning & 0.604  & 0.373  & 0.478   & 0.288  & 0.595 & 0.299 & 0.447            \\
\textbf{ProtChatGPT (Vicuna-13B)}    & \textbf{0.618}  & \textbf{0.399}  & \textbf{0.493}   & \textbf{0.299}  & \textbf{0.640} & \textbf{0.321} & \textbf{0.465}         \\   
\textbf{ProtChatGPT (LL{\scalebox{0.75}[0.75]{A}}MA3-8B)}    & \textbf{0.643}  & \textbf{0.417}  & \textbf{0.515}   & \textbf{0.326}  & \textbf{0.687} & \textbf{0.346} & \textbf{0.487}           
\\ \hline
\end{tabular}
\end{table*}

\begin{figure*}[t]
	\centering
	\includegraphics[width=0.99\textwidth]{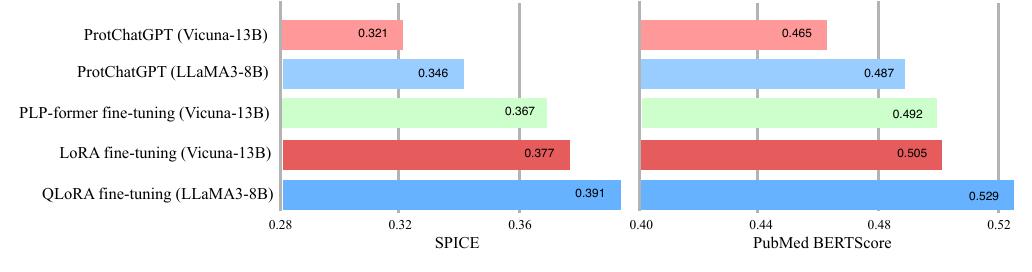}
	\vspace{-2mm}
	\caption{Comparison of fine-tuning of PLP-former and LLM decoders during the instruction tuning stage. We compute the SPICE and PubMed BERTScore for semantic evaluation.}
        \label{fig:ex2}
\end{figure*}

\subsubsection{Ablation Study}
To validate the effectiveness of our method, we further conduct several ablation studies.
To fully showcase the capability of ProtChatGPT, we randomly select 1,000 protein protein-instruction pairs from our refined RCSB-PDB dataset to serve as the test set. 
Note that these testing protein pairs are not used during training for a fair comparison.
We employ seven commonly used metrics in the image captioning and NLP domains to test the performance of ProtChatGPT. 
Detailed descriptions of these metrics can be found in Appendix~\ref{metrics}.

We first devise four variants to validate our contribution. 
(1) \textit{w/o} tertiary structure: remove the branch of the tertiary structure encoding, relying solely on sequence and secondary structure embedding. 
(2) \textit{w/o} secondary structure\&PCG: remove the secondary structure encoder and the Protein Context Gating (PCG) module, directly use sequence and tertiary structure embeddings for contrastive learning. 
(3) \textit{w/o} PLP-former: remove the PLP-former and directly use the sequence embedding from ESM-1b without the injection of protein descriptions.
(4) \textit{w/o} instruction tuning: using the original single descriptions from the RCSB-PDB dataset without generating multiple instruction pairs.
It can be observed from Table~\ref{tab:ex1} that all levels of protein structure information as well as the protein-language alignment play indispensable roles in supplementation and alignment, respectively. 
Notably, in contrast to common metrics like BLEU and METEOR, SPICE and BERTScore pay more attention to deeper semantic information, rather than just lexical and syntactic alignment. 
Particularly, we replace the original Bert encoder with PubMedBERT~\citep{pubmedbert}, a biomedical description-specific encoder pre-trained on large-scale datasets. 
To some extent, this indicator can reflect the scientific validity of the generated responses in the biomedicine domain.

Furthermore, we modified the training strategy of PLP and LLMs in an attempt to achieve better protein-specific dialogue capabilities.
As shown in Figure~\ref{fig:ex2}, we further independently fine-tune the PLP-former and LLM decoders (using 16-bit LoRA~\citep{lora} for Vicuna-13B and 4-bit QLoRA~\citep{qlora} for LL{\scalebox{0.75}[0.75]{A}}MA3-8B) with our refined dataset with instruction pairs.
The performance on two high-level semantic metrics SPICE and PubMed BERTScore indicate that further fine-tuning of both LLMs and PLP-former enhances the performance.
More results under different LLMs and fine-tuning settings are given in Appendix~\ref{moreex}.

\section{Conclusion}
In this paper, we introduce ProtChatGPT, an AI-based protein chat system, to implement ChatGPT-like functionalities for the protein research field.
ProtChatGPT marks the initial effort at bootstrapping multi-level Protein-Language Pretraining (PLP) from pre-trained LLMs for various protein-language tasks.
PLP sits at the intersection between protein and language, which effectively and efficiently enables ProtChatGPT to harvest the off-the-shelf large models from both protein and natural language communities.
Experiments suggest that ProtChatGPT holds potential for application in protein understanding and design. 
We hope this work can facilitate protein research and further inspire other scientific disciplines.

\clearpage

\bibliographystyle{ACM-Reference-Format}
\bibliography{ACM}


\begin{thebibliography}{71}


\ifx \showCODEN    \undefined \def \showCODEN     #1{\unskip}     \fi
\ifx \showDOI      \undefined \def \showDOI       #1{#1}\fi
\ifx \showISBNx    \undefined \def \showISBNx     #1{\unskip}     \fi
\ifx \showISBNxiii \undefined \def \showISBNxiii  #1{\unskip}     \fi
\ifx \showISSN     \undefined \def \showISSN      #1{\unskip}     \fi
\ifx \showLCCN     \undefined \def \showLCCN      #1{\unskip}     \fi
\ifx \shownote     \undefined \def \shownote      #1{#1}          \fi
\ifx \showarticletitle \undefined \def \showarticletitle #1{#1}   \fi
\ifx \showURL      \undefined \def \showURL       {\relax}        \fi
\providecommand\bibfield[2]{#2}
\providecommand\bibinfo[2]{#2}
\providecommand\natexlab[1]{#1}
\providecommand\showeprint[2][]{arXiv:#2}

\bibitem[Alayrac et~al\mbox{.}(2022)]%
        {Flamingo}
\bibfield{author}{\bibinfo{person}{Jean-Baptiste Alayrac}, \bibinfo{person}{Jeff Donahue}, \bibinfo{person}{Pauline Luc}, \bibinfo{person}{Antoine Miech}, \bibinfo{person}{Iain Barr}, \bibinfo{person}{Yana Hasson}, \bibinfo{person}{Karel Lenc}, \bibinfo{person}{Arthur Mensch}, \bibinfo{person}{Katherine Millican}, \bibinfo{person}{Malcolm Reynolds}, {et~al\mbox{.}}} \bibinfo{year}{2022}\natexlab{}.
\newblock \showarticletitle{Flamingo: a visual language model for few-shot learning}.
\newblock \bibinfo{journal}{\emph{Advances in Neural Information Processing Systems}}  \bibinfo{volume}{35} (\bibinfo{year}{2022}), \bibinfo{pages}{23716--23736}.
\newblock


\bibitem[Anderson et~al\mbox{.}(2016)]%
        {spice2016}
\bibfield{author}{\bibinfo{person}{Peter Anderson}, \bibinfo{person}{Basura Fernando}, \bibinfo{person}{Mark Johnson}, {and} \bibinfo{person}{Stephen Gould}.} \bibinfo{year}{2016}\natexlab{}.
\newblock \showarticletitle{SPICE: Semantic Propositional Image Caption Evaluation}. In \bibinfo{booktitle}{\emph{ECCV}}.
\newblock


\bibitem[Banerjee and Lavie(2005)]%
        {banerjee2005meteor}
\bibfield{author}{\bibinfo{person}{Satanjeev Banerjee} {and} \bibinfo{person}{Alon Lavie}.} \bibinfo{year}{2005}\natexlab{}.
\newblock \showarticletitle{METEOR: An automatic metric for MT evaluation with improved correlation with human judgments}. In \bibinfo{booktitle}{\emph{Proceedings of the acl workshop on intrinsic and extrinsic evaluation measures for machine translation and/or summarization}}. \bibinfo{pages}{65--72}.
\newblock


\bibitem[Bao et~al\mbox{.}(2022)]%
        {bao2022vl}
\bibfield{author}{\bibinfo{person}{Hangbo Bao}, \bibinfo{person}{Wenhui Wang}, \bibinfo{person}{Li Dong}, {and} \bibinfo{person}{Furu Wei}.} \bibinfo{year}{2022}\natexlab{}.
\newblock \showarticletitle{Vl-beit: Generative vision-language pretraining}.
\newblock \bibinfo{journal}{\emph{arXiv preprint arXiv:2206.01127}} (\bibinfo{year}{2022}).
\newblock


\bibitem[Berman et~al\mbox{.}(2000)]%
        {PDB}
\bibfield{author}{\bibinfo{person}{Helen~M Berman}, \bibinfo{person}{John Westbrook}, \bibinfo{person}{Zukang Feng}, \bibinfo{person}{Gary Gilliland}, \bibinfo{person}{Talapady~N Bhat}, \bibinfo{person}{Helge Weissig}, \bibinfo{person}{Ilya~N Shindyalov}, {and} \bibinfo{person}{Philip~E Bourne}.} \bibinfo{year}{2000}\natexlab{}.
\newblock \showarticletitle{The protein data bank}.
\newblock \bibinfo{journal}{\emph{Nucleic acids research}} \bibinfo{volume}{28}, \bibinfo{number}{1} (\bibinfo{year}{2000}), \bibinfo{pages}{235--242}.
\newblock


\bibitem[Chen et~al\mbox{.}(2020)]%
        {chen2020uniter}
\bibfield{author}{\bibinfo{person}{Yen-Chun Chen}, \bibinfo{person}{Linjie Li}, \bibinfo{person}{Licheng Yu}, \bibinfo{person}{Ahmed El~Kholy}, \bibinfo{person}{Faisal Ahmed}, \bibinfo{person}{Zhe Gan}, \bibinfo{person}{Yu Cheng}, {and} \bibinfo{person}{Jingjing Liu}.} \bibinfo{year}{2020}\natexlab{}.
\newblock \showarticletitle{{UNITER}: Universal image-text representation learning}. In \bibinfo{booktitle}{\emph{European Conference on Computer Vision (ECCV)}}.
\newblock
\urldef\tempurl%
\url{https://arxiv.org/pdf/1909.11740}
\showURL{%
\tempurl}


\bibitem[Chiang et~al\mbox{.}(2023)]%
        {vicuna}
\bibfield{author}{\bibinfo{person}{Wei-Lin Chiang}, \bibinfo{person}{Zhuohan Li}, \bibinfo{person}{Zi Lin}, \bibinfo{person}{Ying Sheng}, \bibinfo{person}{Zhanghao Wu}, \bibinfo{person}{Hao Zhang}, \bibinfo{person}{Lianmin Zheng}, \bibinfo{person}{Siyuan Zhuang}, \bibinfo{person}{Yonghao Zhuang}, \bibinfo{person}{Joseph~E. Gonzalez}, \bibinfo{person}{Ion Stoica}, {and} \bibinfo{person}{Eric~P. Xing}.} \bibinfo{year}{2023}\natexlab{}.
\newblock \bibinfo{title}{Vicuna: An Open-Source Chatbot Impressing GPT-4 with 90\%* ChatGPT Quality}.
\newblock
\newblock
\urldef\tempurl%
\url{https://lmsys.org/blog/2023-03-30-vicuna/}
\showURL{%
\tempurl}


\bibitem[Chowdhery et~al\mbox{.}(2022)]%
        {palm}
\bibfield{author}{\bibinfo{person}{Aakanksha Chowdhery}, \bibinfo{person}{Sharan Narang}, \bibinfo{person}{Jacob Devlin}, \bibinfo{person}{Maarten Bosma}, \bibinfo{person}{Gaurav Mishra}, \bibinfo{person}{Adam Roberts}, \bibinfo{person}{Paul Barham}, \bibinfo{person}{Hyung~Won Chung}, \bibinfo{person}{Charles Sutton}, \bibinfo{person}{Sebastian Gehrmann}, {et~al\mbox{.}}} \bibinfo{year}{2022}\natexlab{}.
\newblock \showarticletitle{Palm: Scaling language modeling with pathways}.
\newblock \bibinfo{journal}{\emph{arXiv preprint arXiv:2204.02311}} (\bibinfo{year}{2022}).
\newblock


\bibitem[Dai et~al\mbox{.}(2023)]%
        {InstructBLIP}
\bibfield{author}{\bibinfo{person}{Wenliang Dai}, \bibinfo{person}{Junnan Li}, \bibinfo{person}{Dongxu Li}, \bibinfo{person}{Anthony Meng~Huat Tiong}, \bibinfo{person}{Junqi Zhao}, \bibinfo{person}{Weisheng Wang}, \bibinfo{person}{Boyang Li}, \bibinfo{person}{Pascale Fung}, {and} \bibinfo{person}{Steven Hoi}.} \bibinfo{year}{2023}\natexlab{}.
\newblock \bibinfo{title}{InstructBLIP: Towards General-purpose Vision-Language Models with Instruction Tuning}.
\newblock
\newblock
\showeprint[arxiv]{2305.06500}~[cs.CV]


\bibitem[Dettmers et~al\mbox{.}(2024)]%
        {qlora}
\bibfield{author}{\bibinfo{person}{Tim Dettmers}, \bibinfo{person}{Artidoro Pagnoni}, \bibinfo{person}{Ari Holtzman}, {and} \bibinfo{person}{Luke Zettlemoyer}.} \bibinfo{year}{2024}\natexlab{}.
\newblock \showarticletitle{Qlora: Efficient finetuning of quantized llms}.
\newblock \bibinfo{journal}{\emph{Advances in Neural Information Processing Systems}}  \bibinfo{volume}{36} (\bibinfo{year}{2024}).
\newblock


\bibitem[Devlin et~al\mbox{.}(2018)]%
        {bert}
\bibfield{author}{\bibinfo{person}{Jacob Devlin}, \bibinfo{person}{Ming-Wei Chang}, \bibinfo{person}{Kenton Lee}, {and} \bibinfo{person}{Kristina Toutanova}.} \bibinfo{year}{2018}\natexlab{}.
\newblock \showarticletitle{Bert: Pre-training of deep bidirectional transformers for language understanding}.
\newblock \bibinfo{journal}{\emph{arXiv preprint arXiv:1810.04805}} (\bibinfo{year}{2018}).
\newblock


\bibitem[Gligorijevi{\'c} et~al\mbox{.}(2021)]%
        {gligorijevic2021structure}
\bibfield{author}{\bibinfo{person}{Vladimir Gligorijevi{\'c}}, \bibinfo{person}{P~Douglas Renfrew}, \bibinfo{person}{Tomasz Kosciolek}, \bibinfo{person}{Julia~Koehler Leman}, \bibinfo{person}{Daniel Berenberg}, \bibinfo{person}{Tommi Vatanen}, \bibinfo{person}{Chris Chandler}, \bibinfo{person}{Bryn~C Taylor}, \bibinfo{person}{Ian~M Fisk}, \bibinfo{person}{Hera Vlamakis}, {et~al\mbox{.}}} \bibinfo{year}{2021}\natexlab{}.
\newblock \showarticletitle{Structure-based protein function prediction using graph convolutional networks}.
\newblock \bibinfo{journal}{\emph{Nature communications}} \bibinfo{volume}{12}, \bibinfo{number}{1} (\bibinfo{year}{2021}), \bibinfo{pages}{3168}.
\newblock


\bibitem[Gu et~al\mbox{.}(2021)]%
        {pubmedbert}
\bibfield{author}{\bibinfo{person}{Yu Gu}, \bibinfo{person}{Robert Tinn}, \bibinfo{person}{Hao Cheng}, \bibinfo{person}{Michael Lucas}, \bibinfo{person}{Naoto Usuyama}, \bibinfo{person}{Xiaodong Liu}, \bibinfo{person}{Tristan Naumann}, \bibinfo{person}{Jianfeng Gao}, {and} \bibinfo{person}{Hoifung Poon}.} \bibinfo{year}{2021}\natexlab{}.
\newblock \showarticletitle{Domain-specific language model pretraining for biomedical natural language processing}.
\newblock \bibinfo{journal}{\emph{ACM Transactions on Computing for Healthcare (HEALTH)}} \bibinfo{volume}{3}, \bibinfo{number}{1} (\bibinfo{year}{2021}), \bibinfo{pages}{1--23}.
\newblock


\bibitem[Guo et~al\mbox{.}(2023)]%
        {ProteinChat}
\bibfield{author}{\bibinfo{person}{Han Guo}, \bibinfo{person}{Mingjia Huo}, \bibinfo{person}{Ruiyi Zhang}, {and} \bibinfo{person}{Pengtao Xie}.} \bibinfo{year}{2023}\natexlab{}.
\newblock \showarticletitle{Proteinchat: Towards achieving chatgpt-like functionalities on protein 3d structures}.
\newblock \bibinfo{journal}{\emph{Authorea Preprints}} (\bibinfo{year}{2023}).
\newblock


\bibitem[Hsu et~al\mbox{.}(2022)]%
        {ESM-IF}
\bibfield{author}{\bibinfo{person}{Chloe Hsu}, \bibinfo{person}{Robert Verkuil}, \bibinfo{person}{Jason Liu}, \bibinfo{person}{Zeming Lin}, \bibinfo{person}{Brian Hie}, \bibinfo{person}{Tom Sercu}, \bibinfo{person}{Adam Lerer}, {and} \bibinfo{person}{Alexander Rives}.} \bibinfo{year}{2022}\natexlab{}.
\newblock \showarticletitle{Learning inverse folding from millions of predicted structures}. In \bibinfo{booktitle}{\emph{International Conference on Machine Learning}}. \bibinfo{pages}{8946--8970}.
\newblock


\bibitem[Hu et~al\mbox{.}(2021)]%
        {lora}
\bibfield{author}{\bibinfo{person}{Edward~J Hu}, \bibinfo{person}{Yelong Shen}, \bibinfo{person}{Phillip Wallis}, \bibinfo{person}{Zeyuan Allen-Zhu}, \bibinfo{person}{Yuanzhi Li}, \bibinfo{person}{Shean Wang}, \bibinfo{person}{Lu Wang}, {and} \bibinfo{person}{Weizhu Chen}.} \bibinfo{year}{2021}\natexlab{}.
\newblock \showarticletitle{Lora: Low-rank adaptation of large language models}.
\newblock \bibinfo{journal}{\emph{arXiv preprint arXiv:2106.09685}} (\bibinfo{year}{2021}).
\newblock


\bibitem[Jia et~al\mbox{.}(2021)]%
        {jia2021scaling}
\bibfield{author}{\bibinfo{person}{Chao Jia}, \bibinfo{person}{Yinfei Yang}, \bibinfo{person}{Ye Xia}, \bibinfo{person}{Yi-Ting Chen}, \bibinfo{person}{Zarana Parekh}, \bibinfo{person}{Hieu Pham}, \bibinfo{person}{Quoc~V Le}, \bibinfo{person}{Yunhsuan Sung}, \bibinfo{person}{Zhen Li}, {and} \bibinfo{person}{Tom Duerig}.} \bibinfo{year}{2021}\natexlab{}.
\newblock \showarticletitle{Scaling up visual and vision-language representation learning with noisy text supervision}.
\newblock \bibinfo{journal}{\emph{arXiv preprint}} (\bibinfo{year}{2021}).
\newblock


\bibitem[Jin et~al\mbox{.}(2021)]%
        {jin2021iterative}
\bibfield{author}{\bibinfo{person}{Wengong Jin}, \bibinfo{person}{Jeremy Wohlwend}, \bibinfo{person}{Regina Barzilay}, {and} \bibinfo{person}{Tommi Jaakkola}.} \bibinfo{year}{2021}\natexlab{}.
\newblock \showarticletitle{Iterative refinement graph neural network for antibody sequence-structure co-design}.
\newblock \bibinfo{journal}{\emph{arXiv preprint arXiv:2110.04624}} (\bibinfo{year}{2021}).
\newblock


\bibitem[Jing et~al\mbox{.}(2020)]%
        {GPT-4}
\bibfield{author}{\bibinfo{person}{Bowen Jing}, \bibinfo{person}{Stephan Eismann}, \bibinfo{person}{Patricia Suriana}, \bibinfo{person}{Raphael~JL Townshend}, {and} \bibinfo{person}{Ron Dror}.} \bibinfo{year}{2020}\natexlab{}.
\newblock \showarticletitle{Learning from protein structure with geometric vector perceptrons}.
\newblock \bibinfo{journal}{\emph{arXiv preprint arXiv:2009.01411}} (\bibinfo{year}{2020}).
\newblock


\bibitem[Jumper et~al\mbox{.}(2021)]%
        {AlphaFold}
\bibfield{author}{\bibinfo{person}{John Jumper}, \bibinfo{person}{Richard Evans}, \bibinfo{person}{Alexander Pritzel}, \bibinfo{person}{Tim Green}, \bibinfo{person}{Michael Figurnov}, \bibinfo{person}{Olaf Ronneberger}, \bibinfo{person}{Kathryn Tunyasuvunakool}, \bibinfo{person}{Russ Bates}, \bibinfo{person}{Augustin {\v{Z}}{\'\i}dek}, \bibinfo{person}{Anna Potapenko}, {et~al\mbox{.}}} \bibinfo{year}{2021}\natexlab{}.
\newblock \showarticletitle{Highly accurate protein structure prediction with AlphaFold}.
\newblock \bibinfo{journal}{\emph{Nature}} \bibinfo{volume}{596}, \bibinfo{number}{7873} (\bibinfo{year}{2021}), \bibinfo{pages}{583--589}.
\newblock


\bibitem[Kabsch and Sander(1983)]%
        {DSSP}
\bibfield{author}{\bibinfo{person}{Wolfgang Kabsch} {and} \bibinfo{person}{Christian Sander}.} \bibinfo{year}{1983}\natexlab{}.
\newblock \showarticletitle{Dictionary of protein secondary structure: pattern recognition of hydrogen-bonded and geometrical features}.
\newblock \bibinfo{journal}{\emph{Biopolymers: Original Research on Biomolecules}} \bibinfo{volume}{22}, \bibinfo{number}{12} (\bibinfo{year}{1983}), \bibinfo{pages}{2577--2637}.
\newblock


\bibitem[Klausen et~al\mbox{.}(2019)]%
        {NetSurfP}
\bibfield{author}{\bibinfo{person}{Michael~Schantz Klausen}, \bibinfo{person}{Martin~Closter Jespersen}, \bibinfo{person}{Henrik Nielsen}, \bibinfo{person}{Kamilla~Kjaergaard Jensen}, \bibinfo{person}{Vanessa~Isabell Jurtz}, \bibinfo{person}{Casper~Kaae Soenderby}, \bibinfo{person}{Morten Otto~Alexander Sommer}, \bibinfo{person}{Ole Winther}, \bibinfo{person}{Morten Nielsen}, \bibinfo{person}{Bent Petersen}, {et~al\mbox{.}}} \bibinfo{year}{2019}\natexlab{}.
\newblock \showarticletitle{NetSurfP-2.0: Improved prediction of protein structural features by integrated deep learning}.
\newblock \bibinfo{journal}{\emph{Proteins: Structure, Function, and Bioinformatics}} \bibinfo{volume}{87}, \bibinfo{number}{6} (\bibinfo{year}{2019}), \bibinfo{pages}{520--527}.
\newblock


\bibitem[Koh et~al\mbox{.}(2023)]%
        {FROMAGe}
\bibfield{author}{\bibinfo{person}{Jing~Yu Koh}, \bibinfo{person}{Ruslan Salakhutdinov}, {and} \bibinfo{person}{Daniel Fried}.} \bibinfo{year}{2023}\natexlab{}.
\newblock \showarticletitle{Grounding language models to images for multimodal generation}.
\newblock \bibinfo{journal}{\emph{arXiv preprint arXiv:2301.13823}} (\bibinfo{year}{2023}).
\newblock


\bibitem[Kong et~al\mbox{.}(2022)]%
        {kong2022conditional}
\bibfield{author}{\bibinfo{person}{Xiangzhe Kong}, \bibinfo{person}{Wenbing Huang}, {and} \bibinfo{person}{Yang Liu}.} \bibinfo{year}{2022}\natexlab{}.
\newblock \showarticletitle{Conditional antibody design as 3d equivariant graph translation}.
\newblock \bibinfo{journal}{\emph{arXiv preprint arXiv:2208.06073}} (\bibinfo{year}{2022}).
\newblock


\bibitem[Li et~al\mbox{.}(2023)]%
        {BLIP2}
\bibfield{author}{\bibinfo{person}{Junnan Li}, \bibinfo{person}{Dongxu Li}, \bibinfo{person}{Silvio Savarese}, {and} \bibinfo{person}{Steven Hoi}.} \bibinfo{year}{2023}\natexlab{}.
\newblock \showarticletitle{Blip-2: Bootstrapping language-image pre-training with frozen image encoders and large language models}.
\newblock \bibinfo{journal}{\emph{arXiv preprint arXiv:2301.12597}} (\bibinfo{year}{2023}).
\newblock


\bibitem[Li et~al\mbox{.}(2022a)]%
        {BLIP}
\bibfield{author}{\bibinfo{person}{Junnan Li}, \bibinfo{person}{Dongxu Li}, \bibinfo{person}{Caiming Xiong}, {and} \bibinfo{person}{Steven Hoi}.} \bibinfo{year}{2022}\natexlab{a}.
\newblock \showarticletitle{Blip: Bootstrapping language-image pre-training for unified vision-language understanding and generation}. In \bibinfo{booktitle}{\emph{International Conference on Machine Learning}}. \bibinfo{pages}{12888--12900}.
\newblock


\bibitem[Li et~al\mbox{.}(2021)]%
        {li2021align}
\bibfield{author}{\bibinfo{person}{Junnan Li}, \bibinfo{person}{Ramprasaath~R Selvaraju}, \bibinfo{person}{Akhilesh~Deepak Gotmare}, \bibinfo{person}{Shafiq Joty}, \bibinfo{person}{Caiming Xiong}, {and} \bibinfo{person}{Steven Hoi}.} \bibinfo{year}{2021}\natexlab{}.
\newblock \showarticletitle{Align before Fuse: Vision and Language Representation Learning with Momentum Distillation}. In \bibinfo{booktitle}{\emph{Conference on Neural Information Processing Systems (NeurIPS)}}.
\newblock


\bibitem[Li et~al\mbox{.}(2019)]%
        {li2019visualbert}
\bibfield{author}{\bibinfo{person}{Liunian~Harold Li}, \bibinfo{person}{Mark Yatskar}, \bibinfo{person}{Da Yin}, \bibinfo{person}{Cho-Jui Hsieh}, {and} \bibinfo{person}{Kai-Wei Chang}.} \bibinfo{year}{2019}\natexlab{}.
\newblock \showarticletitle{Visual{BERT}: A simple and performant baseline for vision and language}.
\newblock \bibinfo{journal}{\emph{arXiv preprint}} (\bibinfo{year}{2019}).
\newblock
\urldef\tempurl%
\url{https://arxiv.org/pdf/1908.03557}
\showURL{%
\tempurl}


\bibitem[Li et~al\mbox{.}(2022b)]%
        {li2022grounded}
\bibfield{author}{\bibinfo{person}{Liunian~Harold Li}, \bibinfo{person}{Pengchuan Zhang}, \bibinfo{person}{Haotian Zhang}, \bibinfo{person}{Jianwei Yang}, \bibinfo{person}{Chunyuan Li}, \bibinfo{person}{Yiwu Zhong}, \bibinfo{person}{Lijuan Wang}, \bibinfo{person}{Lu Yuan}, \bibinfo{person}{Lei Zhang}, \bibinfo{person}{Jenq-Neng Hwang}, {et~al\mbox{.}}} \bibinfo{year}{2022}\natexlab{b}.
\newblock \showarticletitle{Grounded language-image pre-training}. In \bibinfo{booktitle}{\emph{Proceedings of the IEEE/CVF Conference on Computer Vision and Pattern Recognition}}. \bibinfo{pages}{10965--10975}.
\newblock


\bibitem[Li et~al\mbox{.}(2020)]%
        {li2020oscar}
\bibfield{author}{\bibinfo{person}{Xiujun Li}, \bibinfo{person}{Xi Yin}, \bibinfo{person}{Chunyuan Li}, \bibinfo{person}{Pengchuan Zhang}, \bibinfo{person}{Xiaowei Hu}, \bibinfo{person}{Lei Zhang}, \bibinfo{person}{Lijuan Wang}, \bibinfo{person}{Houdong Hu}, \bibinfo{person}{Li Dong}, \bibinfo{person}{Furu Wei}, {et~al\mbox{.}}} \bibinfo{year}{2020}\natexlab{}.
\newblock \showarticletitle{Oscar: Object-semantics aligned pre-training for vision-language tasks}. In \bibinfo{booktitle}{\emph{European Conference on Computer Vision (ECCV)}}.
\newblock
\urldef\tempurl%
\url{https://arxiv.org/pdf/2004.06165}
\showURL{%
\tempurl}


\bibitem[Lin and Hovy(2002)]%
        {lin2002manual}
\bibfield{author}{\bibinfo{person}{Chin-Yew Lin} {and} \bibinfo{person}{Eduard Hovy}.} \bibinfo{year}{2002}\natexlab{}.
\newblock \showarticletitle{Manual and automatic evaluation of summaries}. In \bibinfo{booktitle}{\emph{Proceedings of the ACL-02 workshop on automatic summarization}}. \bibinfo{pages}{45--51}.
\newblock


\bibitem[Lin et~al\mbox{.}(2023)]%
        {ESM-2}
\bibfield{author}{\bibinfo{person}{Zeming Lin}, \bibinfo{person}{Halil Akin}, \bibinfo{person}{Roshan Rao}, \bibinfo{person}{Brian Hie}, \bibinfo{person}{Zhongkai Zhu}, \bibinfo{person}{Wenting Lu}, \bibinfo{person}{Nikita Smetanin}, \bibinfo{person}{Robert Verkuil}, \bibinfo{person}{Ori Kabeli}, \bibinfo{person}{Yaniv Shmueli}, {et~al\mbox{.}}} \bibinfo{year}{2023}\natexlab{}.
\newblock \showarticletitle{Evolutionary-scale prediction of atomic-level protein structure with a language model}.
\newblock \bibinfo{journal}{\emph{Science}} \bibinfo{volume}{379}, \bibinfo{number}{6637} (\bibinfo{year}{2023}), \bibinfo{pages}{1123--1130}.
\newblock


\bibitem[Liu et~al\mbox{.}(2023)]%
        {liu2023text}
\bibfield{author}{\bibinfo{person}{Shengchao Liu}, \bibinfo{person}{Yanjing Li}, \bibinfo{person}{Zhuoxinran Li}, \bibinfo{person}{Anthony Gitter}, \bibinfo{person}{Yutao Zhu}, \bibinfo{person}{Jiarui Lu}, \bibinfo{person}{Zhao Xu}, \bibinfo{person}{Weili Nie}, \bibinfo{person}{Arvind Ramanathan}, \bibinfo{person}{Chaowei Xiao}, {et~al\mbox{.}}} \bibinfo{year}{2023}\natexlab{}.
\newblock \showarticletitle{A text-guided protein design framework}.
\newblock \bibinfo{journal}{\emph{arXiv preprint arXiv:2302.04611}} (\bibinfo{year}{2023}).
\newblock


\bibitem[Liu et~al\mbox{.}(2024)]%
        {ProtT3}
\bibfield{author}{\bibinfo{person}{Zhiyuan Liu}, \bibinfo{person}{An Zhang}, \bibinfo{person}{Hao Fei}, \bibinfo{person}{Enzhi Zhang}, \bibinfo{person}{Xiang Wang}, \bibinfo{person}{Kenji Kawaguchi}, {and} \bibinfo{person}{Tat-Seng Chua}.} \bibinfo{year}{2024}\natexlab{}.
\newblock \showarticletitle{ProtT3: Protein-to-Text Generation for Text-based Protein Understanding}.
\newblock \bibinfo{journal}{\emph{arXiv preprint arXiv:2405.12564}} (\bibinfo{year}{2024}).
\newblock


\bibitem[Loshchilov and Hutter(2017)]%
        {adamw}
\bibfield{author}{\bibinfo{person}{Ilya Loshchilov} {and} \bibinfo{person}{Frank Hutter}.} \bibinfo{year}{2017}\natexlab{}.
\newblock \showarticletitle{Decoupled weight decay regularization}.
\newblock \bibinfo{journal}{\emph{arXiv preprint arXiv:1711.05101}} (\bibinfo{year}{2017}).
\newblock


\bibitem[Lu et~al\mbox{.}(2019)]%
        {lu2019vilbert}
\bibfield{author}{\bibinfo{person}{Jiasen Lu}, \bibinfo{person}{Dhruv Batra}, \bibinfo{person}{Devi Parikh}, {and} \bibinfo{person}{Stefan Lee}.} \bibinfo{year}{2019}\natexlab{}.
\newblock \showarticletitle{Vilbert: Pretraining task-agnostic visiolinguistic representations for vision-and-language tasks}. In \bibinfo{booktitle}{\emph{Conference on Neural Information Processing Systems (NeurIPS)}}.
\newblock


\bibitem[Madani et~al\mbox{.}(2023)]%
        {madani2023large}
\bibfield{author}{\bibinfo{person}{Ali Madani}, \bibinfo{person}{Ben Krause}, \bibinfo{person}{Eric~R Greene}, \bibinfo{person}{Subu Subramanian}, \bibinfo{person}{Benjamin~P Mohr}, \bibinfo{person}{James~M Holton}, \bibinfo{person}{Jose~Luis Olmos~Jr}, \bibinfo{person}{Caiming Xiong}, \bibinfo{person}{Zachary~Z Sun}, \bibinfo{person}{Richard Socher}, {et~al\mbox{.}}} \bibinfo{year}{2023}\natexlab{}.
\newblock \showarticletitle{Large language models generate functional protein sequences across diverse families}.
\newblock \bibinfo{journal}{\emph{Nature Biotechnology}} (\bibinfo{year}{2023}), \bibinfo{pages}{1--8}.
\newblock


\bibitem[Mahbub and Bayzid(2022)]%
        {mahbub2022egret}
\bibfield{author}{\bibinfo{person}{Sazan Mahbub} {and} \bibinfo{person}{Md~Shamsuzzoha Bayzid}.} \bibinfo{year}{2022}\natexlab{}.
\newblock \showarticletitle{EGRET: edge aggregated graph attention networks and transfer learning improve protein--protein interaction site prediction}.
\newblock \bibinfo{journal}{\emph{Briefings in Bioinformatics}} \bibinfo{volume}{23}, \bibinfo{number}{2} (\bibinfo{year}{2022}), \bibinfo{pages}{bbab578}.
\newblock


\bibitem[Meier et~al\mbox{.}(2021)]%
        {ESM-1v}
\bibfield{author}{\bibinfo{person}{Joshua Meier}, \bibinfo{person}{Roshan Rao}, \bibinfo{person}{Robert Verkuil}, \bibinfo{person}{Jason Liu}, \bibinfo{person}{Tom Sercu}, {and} \bibinfo{person}{Alex Rives}.} \bibinfo{year}{2021}\natexlab{}.
\newblock \showarticletitle{Language models enable zero-shot prediction of the effects of mutations on protein function}.
\newblock \bibinfo{journal}{\emph{Advances in Neural Information Processing Systems}}  \bibinfo{volume}{34} (\bibinfo{year}{2021}), \bibinfo{pages}{29287--29303}.
\newblock


\bibitem[Mistry et~al\mbox{.}(2021)]%
        {pfam}
\bibfield{author}{\bibinfo{person}{Jaina Mistry}, \bibinfo{person}{Sara Chuguransky}, \bibinfo{person}{Lowri Williams}, \bibinfo{person}{Matloob Qureshi}, \bibinfo{person}{Gustavo~A Salazar}, \bibinfo{person}{Erik~LL Sonnhammer}, \bibinfo{person}{Silvio~CE Tosatto}, \bibinfo{person}{Lisanna Paladin}, \bibinfo{person}{Shriya Raj}, \bibinfo{person}{Lorna~J Richardson}, {et~al\mbox{.}}} \bibinfo{year}{2021}\natexlab{}.
\newblock \showarticletitle{Pfam: The protein families database in 2021}.
\newblock \bibinfo{journal}{\emph{Nucleic acids research}} \bibinfo{volume}{49}, \bibinfo{number}{D1} (\bibinfo{year}{2021}), \bibinfo{pages}{D412--D419}.
\newblock


\bibitem[Nayak et~al\mbox{.}(2024)]%
        {Bonito}
\bibfield{author}{\bibinfo{person}{Nihal~V Nayak}, \bibinfo{person}{Yiyang Nan}, \bibinfo{person}{Avi Trost}, {and} \bibinfo{person}{Stephen~H Bach}.} \bibinfo{year}{2024}\natexlab{}.
\newblock \showarticletitle{Learning to Generate Instruction Tuning Datasets for Zero-Shot Task Adaptation}.
\newblock \bibinfo{journal}{\emph{arXiv preprint arXiv:2402.18334}} (\bibinfo{year}{2024}).
\newblock


\bibitem[Notin et~al\mbox{.}(2022)]%
        {notin2022tranception}
\bibfield{author}{\bibinfo{person}{Pascal Notin}, \bibinfo{person}{Mafalda Dias}, \bibinfo{person}{Jonathan Frazer}, \bibinfo{person}{Javier~Marchena Hurtado}, \bibinfo{person}{Aidan~N Gomez}, \bibinfo{person}{Debora Marks}, {and} \bibinfo{person}{Yarin Gal}.} \bibinfo{year}{2022}\natexlab{}.
\newblock \showarticletitle{Tranception: protein fitness prediction with autoregressive transformers and inference-time retrieval}. In \bibinfo{booktitle}{\emph{International Conference on Machine Learning}}. \bibinfo{pages}{16990--17017}.
\newblock


\bibitem[OpenAI(2023)]%
        {GVP}
\bibfield{author}{\bibinfo{person}{OpenAI}.} \bibinfo{year}{2023}\natexlab{}.
\newblock \showarticletitle{GPT-4 Technical Report}.
\newblock \bibinfo{journal}{\emph{arXiv preprint arXiv:2303.08774}} (\bibinfo{year}{2023}).
\newblock


\bibitem[Papineni et~al\mbox{.}(2002)]%
        {papineni2002bleu}
\bibfield{author}{\bibinfo{person}{Kishore Papineni}, \bibinfo{person}{Salim Roukos}, \bibinfo{person}{Todd Ward}, {and} \bibinfo{person}{Wei-Jing Zhu}.} \bibinfo{year}{2002}\natexlab{}.
\newblock \showarticletitle{Bleu: a method for automatic evaluation of machine translation}. In \bibinfo{booktitle}{\emph{Proceedings of the 40th annual meeting of the Association for Computational Linguistics}}. \bibinfo{pages}{311--318}.
\newblock


\bibitem[Radford et~al\mbox{.}(2021a)]%
        {VLP}
\bibfield{author}{\bibinfo{person}{Alec Radford}, \bibinfo{person}{Jong~Wook Kim}, \bibinfo{person}{Chris Hallacy}, \bibinfo{person}{Aditya Ramesh}, \bibinfo{person}{Gabriel Goh}, \bibinfo{person}{Sandhini Agarwal}, \bibinfo{person}{Girish Sastry}, \bibinfo{person}{Amanda Askell}, \bibinfo{person}{Pamela Mishkin}, \bibinfo{person}{Jack Clark}, {et~al\mbox{.}}} \bibinfo{year}{2021}\natexlab{a}.
\newblock \showarticletitle{Learning transferable visual models from natural language supervision}. In \bibinfo{booktitle}{\emph{International conference on machine learning}}. PMLR, \bibinfo{pages}{8748--8763}.
\newblock


\bibitem[Radford et~al\mbox{.}(2021b)]%
        {radford2021learning}
\bibfield{author}{\bibinfo{person}{Alec Radford}, \bibinfo{person}{Jong~Wook Kim}, \bibinfo{person}{Chris Hallacy}, \bibinfo{person}{Aditya Ramesh}, \bibinfo{person}{Gabriel Goh}, \bibinfo{person}{Sandhini Agarwal}, \bibinfo{person}{Girish Sastry}, \bibinfo{person}{Amanda Askell}, \bibinfo{person}{Pamela Mishkin}, \bibinfo{person}{Jack Clark}, {et~al\mbox{.}}} \bibinfo{year}{2021}\natexlab{b}.
\newblock \showarticletitle{Learning transferable visual models from natural language supervision}. In \bibinfo{booktitle}{\emph{International Conference on Machine Learning (ICML)}}.
\newblock


\bibitem[Radford et~al\mbox{.}(2019)]%
        {GPT}
\bibfield{author}{\bibinfo{person}{Alec Radford}, \bibinfo{person}{Jeffrey Wu}, \bibinfo{person}{Rewon Child}, \bibinfo{person}{David Luan}, \bibinfo{person}{Dario Amodei}, \bibinfo{person}{Ilya Sutskever}, {et~al\mbox{.}}} \bibinfo{year}{2019}\natexlab{}.
\newblock \showarticletitle{Language models are unsupervised multitask learners}.
\newblock \bibinfo{journal}{\emph{OpenAI blog}} \bibinfo{volume}{1}, \bibinfo{number}{8} (\bibinfo{year}{2019}), \bibinfo{pages}{9}.
\newblock


\bibitem[Raffel et~al\mbox{.}(2020)]%
        {T5}
\bibfield{author}{\bibinfo{person}{Colin Raffel}, \bibinfo{person}{Noam Shazeer}, \bibinfo{person}{Adam Roberts}, \bibinfo{person}{Katherine Lee}, \bibinfo{person}{Sharan Narang}, \bibinfo{person}{Michael Matena}, \bibinfo{person}{Yanqi Zhou}, \bibinfo{person}{Wei Li}, {and} \bibinfo{person}{Peter~J Liu}.} \bibinfo{year}{2020}\natexlab{}.
\newblock \showarticletitle{Exploring the limits of transfer learning with a unified text-to-text transformer}.
\newblock \bibinfo{journal}{\emph{The Journal of Machine Learning Research}} \bibinfo{volume}{21}, \bibinfo{number}{1} (\bibinfo{year}{2020}), \bibinfo{pages}{5485--5551}.
\newblock


\bibitem[Rao et~al\mbox{.}(2020)]%
        {ESM}
\bibfield{author}{\bibinfo{person}{Roshan Rao}, \bibinfo{person}{Joshua Meier}, \bibinfo{person}{Tom Sercu}, \bibinfo{person}{Sergey Ovchinnikov}, {and} \bibinfo{person}{Alexander Rives}.} \bibinfo{year}{2020}\natexlab{}.
\newblock \showarticletitle{Transformer protein language models are unsupervised structure learners}.
\newblock \bibinfo{journal}{\emph{Biorxiv}} (\bibinfo{year}{2020}), \bibinfo{pages}{2020--12}.
\newblock


\bibitem[R{\'e}au et~al\mbox{.}(2023)]%
        {reau2023deeprank}
\bibfield{author}{\bibinfo{person}{Manon R{\'e}au}, \bibinfo{person}{Nicolas Renaud}, \bibinfo{person}{Li~C Xue}, {and} \bibinfo{person}{Alexandre~MJJ Bonvin}.} \bibinfo{year}{2023}\natexlab{}.
\newblock \showarticletitle{DeepRank-GNN: a graph neural network framework to learn patterns in protein--protein interfaces}.
\newblock \bibinfo{journal}{\emph{Bioinformatics}} \bibinfo{volume}{39}, \bibinfo{number}{1} (\bibinfo{year}{2023}), \bibinfo{pages}{btac759}.
\newblock


\bibitem[Rives et~al\mbox{.}(2021)]%
        {ESM-1b}
\bibfield{author}{\bibinfo{person}{Alexander Rives}, \bibinfo{person}{Joshua Meier}, \bibinfo{person}{Tom Sercu}, \bibinfo{person}{Siddharth Goyal}, \bibinfo{person}{Zeming Lin}, \bibinfo{person}{Jason Liu}, \bibinfo{person}{Demi Guo}, \bibinfo{person}{Myle Ott}, \bibinfo{person}{C~Lawrence Zitnick}, \bibinfo{person}{Jerry Ma}, {et~al\mbox{.}}} \bibinfo{year}{2021}\natexlab{}.
\newblock \showarticletitle{Biological structure and function emerge from scaling unsupervised learning to 250 million protein sequences}.
\newblock \bibinfo{journal}{\emph{Proceedings of the National Academy of Sciences}} \bibinfo{volume}{118}, \bibinfo{number}{15} (\bibinfo{year}{2021}), \bibinfo{pages}{e2016239118}.
\newblock


\bibitem[Taylor et~al\mbox{.}(2022)]%
        {galactica}
\bibfield{author}{\bibinfo{person}{Ross Taylor}, \bibinfo{person}{Marcin Kardas}, \bibinfo{person}{Guillem Cucurull}, \bibinfo{person}{Thomas Scialom}, \bibinfo{person}{Anthony Hartshorn}, \bibinfo{person}{Elvis Saravia}, \bibinfo{person}{Andrew Poulton}, \bibinfo{person}{Viktor Kerkez}, {and} \bibinfo{person}{Robert Stojnic}.} \bibinfo{year}{2022}\natexlab{}.
\newblock \showarticletitle{Galactica: A large language model for science}.
\newblock \bibinfo{journal}{\emph{arXiv preprint arXiv:2211.09085}} (\bibinfo{year}{2022}).
\newblock


\bibitem[Touvron et~al\mbox{.}(2023)]%
        {llama}
\bibfield{author}{\bibinfo{person}{Hugo Touvron}, \bibinfo{person}{Thibaut Lavril}, \bibinfo{person}{Gautier Izacard}, \bibinfo{person}{Xavier Martinet}, \bibinfo{person}{Marie-Anne Lachaux}, \bibinfo{person}{Timoth{\'e}e Lacroix}, \bibinfo{person}{Baptiste Rozi{\`e}re}, \bibinfo{person}{Naman Goyal}, \bibinfo{person}{Eric Hambro}, \bibinfo{person}{Faisal Azhar}, {et~al\mbox{.}}} \bibinfo{year}{2023}\natexlab{}.
\newblock \showarticletitle{Llama: Open and efficient foundation language models}.
\newblock \bibinfo{journal}{\emph{arXiv preprint arXiv:2302.13971}} (\bibinfo{year}{2023}).
\newblock


\bibitem[van Sonsbeek et~al\mbox{.}(2023)]%
        {MedVQA}
\bibfield{author}{\bibinfo{person}{Tom van Sonsbeek}, \bibinfo{person}{Mohammad~Mahdi Derakhshani}, \bibinfo{person}{Ivona Najdenkoska}, \bibinfo{person}{Cees~GM Snoek}, {and} \bibinfo{person}{Marcel Worring}.} \bibinfo{year}{2023}\natexlab{}.
\newblock \showarticletitle{Open-ended medical visual question answering through prefix tuning of language models}.
\newblock \bibinfo{journal}{\emph{arXiv preprint arXiv:2303.05977}} (\bibinfo{year}{2023}).
\newblock


\bibitem[Varadi et~al\mbox{.}(2022)]%
        {varadi2022alphafold}
\bibfield{author}{\bibinfo{person}{Mihaly Varadi}, \bibinfo{person}{Stephen Anyango}, \bibinfo{person}{Mandar Deshpande}, \bibinfo{person}{Sreenath Nair}, \bibinfo{person}{Cindy Natassia}, \bibinfo{person}{Galabina Yordanova}, \bibinfo{person}{David Yuan}, \bibinfo{person}{Oana Stroe}, \bibinfo{person}{Gemma Wood}, \bibinfo{person}{Agata Laydon}, {et~al\mbox{.}}} \bibinfo{year}{2022}\natexlab{}.
\newblock \showarticletitle{AlphaFold Protein Structure Database: massively expanding the structural coverage of protein-sequence space with high-accuracy models}.
\newblock \bibinfo{journal}{\emph{Nucleic acids research}} \bibinfo{volume}{50}, \bibinfo{number}{D1} (\bibinfo{year}{2022}), \bibinfo{pages}{D439--D444}.
\newblock


\bibitem[Vaswani et~al\mbox{.}(2017)]%
        {transformer}
\bibfield{author}{\bibinfo{person}{Ashish Vaswani}, \bibinfo{person}{Noam Shazeer}, \bibinfo{person}{Niki Parmar}, \bibinfo{person}{Jakob Uszkoreit}, \bibinfo{person}{Llion Jones}, \bibinfo{person}{Aidan~N Gomez}, \bibinfo{person}{{\L}ukasz Kaiser}, {and} \bibinfo{person}{Illia Polosukhin}.} \bibinfo{year}{2017}\natexlab{}.
\newblock \showarticletitle{Attention is all you need}.
\newblock \bibinfo{journal}{\emph{Advances in neural information processing systems}}  \bibinfo{volume}{30} (\bibinfo{year}{2017}).
\newblock


\bibitem[Vedantam et~al\mbox{.}(2015)]%
        {vedantam2015cider}
\bibfield{author}{\bibinfo{person}{Ramakrishna Vedantam}, \bibinfo{person}{C Lawrence~Zitnick}, {and} \bibinfo{person}{Devi Parikh}.} \bibinfo{year}{2015}\natexlab{}.
\newblock \showarticletitle{Cider: Consensus-based image description evaluation}. In \bibinfo{booktitle}{\emph{Proceedings of the IEEE conference on computer vision and pattern recognition}}. \bibinfo{pages}{4566--4575}.
\newblock


\bibitem[Vig et~al\mbox{.}(2020)]%
        {vig2020bertology}
\bibfield{author}{\bibinfo{person}{Jesse Vig}, \bibinfo{person}{Ali Madani}, \bibinfo{person}{Lav~R Varshney}, \bibinfo{person}{Caiming Xiong}, \bibinfo{person}{Richard Socher}, {and} \bibinfo{person}{Nazneen~Fatema Rajani}.} \bibinfo{year}{2020}\natexlab{}.
\newblock \showarticletitle{Bertology meets biology: Interpreting attention in protein language models}.
\newblock \bibinfo{journal}{\emph{arXiv preprint arXiv:2006.15222}} (\bibinfo{year}{2020}).
\newblock


\bibitem[Wang et~al\mbox{.}(2023)]%
        {ChatCAD}
\bibfield{author}{\bibinfo{person}{Sheng Wang}, \bibinfo{person}{Zihao Zhao}, \bibinfo{person}{Xi Ouyang}, \bibinfo{person}{Qian Wang}, {and} \bibinfo{person}{Dinggang Shen}.} \bibinfo{year}{2023}\natexlab{}.
\newblock \showarticletitle{Chatcad: Interactive computer-aided diagnosis on medical image using large language models}.
\newblock \bibinfo{journal}{\emph{arXiv preprint arXiv:2302.07257}} (\bibinfo{year}{2023}).
\newblock


\bibitem[Wang et~al\mbox{.}(2022)]%
        {wang2022vlmixer}
\bibfield{author}{\bibinfo{person}{Teng Wang}, \bibinfo{person}{Wenhao Jiang}, \bibinfo{person}{Zhichao Lu}, \bibinfo{person}{Feng Zheng}, \bibinfo{person}{Ran Cheng}, \bibinfo{person}{Chengguo Yin}, {and} \bibinfo{person}{Ping Luo}.} \bibinfo{year}{2022}\natexlab{}.
\newblock \showarticletitle{VLMixer: Unpaired Vision-Language Pre-training via Cross-Modal CutMix}. In \bibinfo{booktitle}{\emph{International Conference on Machine Learning}}. PMLR, \bibinfo{pages}{22680--22690}.
\newblock


\bibitem[Wang et~al\mbox{.}(2021)]%
        {wang2021simvlm}
\bibfield{author}{\bibinfo{person}{Zirui Wang}, \bibinfo{person}{Jiahui Yu}, \bibinfo{person}{Adams~Wei Yu}, \bibinfo{person}{Zihang Dai}, \bibinfo{person}{Yulia Tsvetkov}, {and} \bibinfo{person}{Yuan Cao}.} \bibinfo{year}{2021}\natexlab{}.
\newblock \showarticletitle{SimVLM: Simple Visual Language Model Pretraining with Weak Supervision}.
\newblock \bibinfo{journal}{\emph{arXiv preprint}} (\bibinfo{year}{2021}).
\newblock


\bibitem[Xu et~al\mbox{.}(2023)]%
        {protst}
\bibfield{author}{\bibinfo{person}{Minghao Xu}, \bibinfo{person}{Xinyu Yuan}, \bibinfo{person}{Santiago Miret}, {and} \bibinfo{person}{Jian Tang}.} \bibinfo{year}{2023}\natexlab{}.
\newblock \showarticletitle{Protst: Multi-modality learning of protein sequences and biomedical texts}.
\newblock \bibinfo{journal}{\emph{arXiv preprint arXiv:2301.12040}} (\bibinfo{year}{2023}).
\newblock


\bibitem[Yao et~al\mbox{.}(2021)]%
        {yao2021filip}
\bibfield{author}{\bibinfo{person}{Lewei Yao}, \bibinfo{person}{Runhui Huang}, \bibinfo{person}{Lu Hou}, \bibinfo{person}{Guansong Lu}, \bibinfo{person}{Minzhe Niu}, \bibinfo{person}{Hang Xu}, \bibinfo{person}{Xiaodan Liang}, \bibinfo{person}{Zhenguo Li}, \bibinfo{person}{Xin Jiang}, {and} \bibinfo{person}{Chunjing Xu}.} \bibinfo{year}{2021}\natexlab{}.
\newblock \showarticletitle{Filip: Fine-grained interactive language-image pre-training}.
\newblock \bibinfo{journal}{\emph{arXiv preprint arXiv:2111.07783}} (\bibinfo{year}{2021}).
\newblock


\bibitem[Yu et~al\mbox{.}(2022)]%
        {yu2022coca}
\bibfield{author}{\bibinfo{person}{Jiahui Yu}, \bibinfo{person}{Zirui Wang}, \bibinfo{person}{Vijay Vasudevan}, \bibinfo{person}{Legg Yeung}, \bibinfo{person}{Mojtaba Seyedhosseini}, {and} \bibinfo{person}{Yonghui Wu}.} \bibinfo{year}{2022}\natexlab{}.
\newblock \showarticletitle{Coca: Contrastive captioners are image-text foundation models}.
\newblock \bibinfo{journal}{\emph{arXiv preprint arXiv:2205.01917}} (\bibinfo{year}{2022}).
\newblock


\bibitem[Zeng et~al\mbox{.}(2021)]%
        {zeng2021multi}
\bibfield{author}{\bibinfo{person}{Yan Zeng}, \bibinfo{person}{Xinsong Zhang}, {and} \bibinfo{person}{Hang Li}.} \bibinfo{year}{2021}\natexlab{}.
\newblock \showarticletitle{Multi-Grained Vision Language Pre-Training: Aligning Texts with Visual Concepts}.
\newblock \bibinfo{journal}{\emph{arXiv preprint arXiv:2111.08276}} (\bibinfo{year}{2021}).
\newblock


\bibitem[Zhang et~al\mbox{.}(2022a)]%
        {zhang2022ontoprotein}
\bibfield{author}{\bibinfo{person}{Ningyu Zhang}, \bibinfo{person}{Zhen Bi}, \bibinfo{person}{Xiaozhuan Liang}, \bibinfo{person}{Siyuan Cheng}, \bibinfo{person}{Haosen Hong}, \bibinfo{person}{Shumin Deng}, \bibinfo{person}{Jiazhang Lian}, \bibinfo{person}{Qiang Zhang}, {and} \bibinfo{person}{Huajun Chen}.} \bibinfo{year}{2022}\natexlab{a}.
\newblock \showarticletitle{Ontoprotein: Protein pretraining with gene ontology embedding}.
\newblock \bibinfo{journal}{\emph{arXiv preprint arXiv:2201.11147}} (\bibinfo{year}{2022}).
\newblock


\bibitem[Zhang et~al\mbox{.}(2022b)]%
        {Ontoprotein}
\bibfield{author}{\bibinfo{person}{Ningyu Zhang}, \bibinfo{person}{Zhen Bi}, \bibinfo{person}{Xiaozhuan Liang}, \bibinfo{person}{Siyuan Cheng}, \bibinfo{person}{Haosen Hong}, \bibinfo{person}{Shumin Deng}, \bibinfo{person}{Jiazhang Lian}, \bibinfo{person}{Qiang Zhang}, {and} \bibinfo{person}{Huajun Chen}.} \bibinfo{year}{2022}\natexlab{b}.
\newblock \showarticletitle{Ontoprotein: Protein pretraining with gene ontology embedding}. In \bibinfo{booktitle}{\emph{International Conference on Learning Representations (ICLR)}}.
\newblock


\bibitem[Zhang et~al\mbox{.}(2021)]%
        {zhang2021vinvl}
\bibfield{author}{\bibinfo{person}{Pengchuan Zhang}, \bibinfo{person}{Xiujun Li}, \bibinfo{person}{Xiaowei Hu}, \bibinfo{person}{Jianwei Yang}, \bibinfo{person}{Lei Zhang}, \bibinfo{person}{Lijuan Wang}, \bibinfo{person}{Yejin Choi}, {and} \bibinfo{person}{Jianfeng Gao}.} \bibinfo{year}{2021}\natexlab{}.
\newblock \showarticletitle{{VinVL}: Revisiting visual representations in vision-language models}. In \bibinfo{booktitle}{\emph{Conference on Computer Vision and Pattern Recognition (CVPR)}}.
\newblock


\bibitem[Zhang et~al\mbox{.}(2019)]%
        {zhang2019bertscore}
\bibfield{author}{\bibinfo{person}{Tianyi Zhang}, \bibinfo{person}{Varsha Kishore}, \bibinfo{person}{Felix Wu}, \bibinfo{person}{Kilian~Q Weinberger}, {and} \bibinfo{person}{Yoav Artzi}.} \bibinfo{year}{2019}\natexlab{}.
\newblock \showarticletitle{Bertscore: Evaluating text generation with bert}.
\newblock \bibinfo{journal}{\emph{arXiv preprint arXiv:1904.09675}} (\bibinfo{year}{2019}).
\newblock


\bibitem[Zhang et~al\mbox{.}(2023)]%
        {GearNet}
\bibfield{author}{\bibinfo{person}{Zuobai Zhang}, \bibinfo{person}{Minghao Xu}, \bibinfo{person}{Arian Jamasb}, \bibinfo{person}{Vijil Chenthamarakshan}, \bibinfo{person}{Aurelie Lozano}, \bibinfo{person}{Payel Das}, {and} \bibinfo{person}{Jian Tang}.} \bibinfo{year}{2023}\natexlab{}.
\newblock \showarticletitle{Protein representation learning by geometric structure pretraining}. In \bibinfo{booktitle}{\emph{International Conference on Learning Representations}}.
\newblock


\bibitem[Zhu et~al\mbox{.}(2023)]%
        {miniGPT4}
\bibfield{author}{\bibinfo{person}{Deyao Zhu}, \bibinfo{person}{Jun Chen}, \bibinfo{person}{Xiaoqian Shen}, \bibinfo{person}{Xiang Li}, {and} \bibinfo{person}{Mohamed Elhoseiny}.} \bibinfo{year}{2023}\natexlab{}.
\newblock \showarticletitle{Minigpt-4: Enhancing vision-language understanding with advanced large language models}.
\newblock \bibinfo{journal}{\emph{arXiv preprint arXiv:2304.10592}} (\bibinfo{year}{2023}).
\newblock


\end{thebibliography}

\clearpage
\section*{Appendix}

Appendix~\ref{limitation}: The limitation and discussion of ProtChatGPT.

Appendix~\ref{broader}: The broader impact of ProtChatGPT.

Appendix~\ref{Proteins}: We provide a brief introduction on proteins, including protein structures, homologous Proteins, mutually Exclusive Functions, and Dictionary of Protein Secondary Structure (DSSP).

Appendix~\ref{VLP}: Introduction and related work on Vision-Language Pretraining (VLP).

Appendix~\ref{PLPdetail}: Details on the proposed Protein-Language Representation Learning (PLP).

Appendix~\ref{appedix:dataset}: Details and examples of the training Datasets.

Appendix~\ref{implementation}: More implementation details.

Appendix~\ref{moreex}: Additional experiments on LLM fine-tuning and protein retrieval.

Appendix~\ref{metrics}: Details on evaluation metrics used in this paper.

\appendix

\section{Limitation and Discussion}\label{limitation}
ProtChatGPT leverages the capabilities of LLMs for protein-specialized conversations. 
However, it inherits LLM’s potential \textit{language hallucination}.
It is an indispensable concern especially when it relates to protein research and healthcare. 
Although the two case studies in Section~\ref{case} demonstrate that our ProtChatGPT can distinguish some complex cases (\eg, homologous proteins and mutually exclusive functions), these capabilities heavily depend on the presence of similar examples in our training dataset, especially the description part.
Given an unknown protein, ProtChatGPT might produce certain descriptions that sound correct but lack proper scientific verification, possibly leading researchers astray. 
This issue might be alleviated by training the model with more high-quality, aligned protein-text pairs, or aligning with more advanced LLMs in the future.
In this manner, rigorous data processing and selection strategies should be implemented to ensure the validity and reliability of the training data. 
Feedback from domain experts is also important to refine the model. 
Combined with reinforcement or continual learning techniques, ProtChatGPT could keep improving the quality of its responses.
Furthermore, training only one projection adapter might not provide enough capacity to learn extensive protein-text alignment. 
This issue could be alleviated by designing a more powerful adapter to facilitate the interactions between sequence and structure embeddings or utilizing other powerful protein encoders such as GearNet~\citep{GearNet} and ESM-2~\citep{ESM-2}. 
In future work, addressing these issues and refining ProtChatGPT is essential.
With ongoing improvements and regular expert feedback, ProtChatGPT has the potential to become a trusted assistant in protein research, offering more valuable insights for further investigations.

\section{Broader Impact}\label{broader}
ProtChatGPT extends across multiple dimensions of protein research and biotechnology. 
By facilitating dialogue-based interactions with a protein-specific model, ProtChatGPT could democratize access to complex protein analysis, enabling researchers, regardless of their computational expertise, to engage deeply with protein data. 
This could accelerate discovery processes in drug design, genetic research, and disease understanding by providing intuitive, conversational access to sophisticated protein analyses and hypothesis generation.
Furthermore, ProtChatGPT could also serve educational purposes, aiding in the training of next-generation scientists by providing an interactive learning tool that offers immediate feedback and explanations about protein structures and functions. 
This might inspire innovative approaches in both academia and industry, potentially leading to novel therapeutic strategies and an enhanced understanding of biological mechanisms.

\section{Brief Introduction on Proteins}\label{Proteins}
Proteins are large molecules formed by amino acids linked through peptide bonds and are essential components of cells and organisms. 
They perform a range of critical functions in the body, from providing structural support to regulating biochemical processes.
Proteins are made up of hundreds or thousands of smaller units called amino acids, which are attached to one another in long chains. 
There are 20 different types of amino acids that can be combined to make a protein. 
The sequence of amino acids determines the unique 3-dimensional structure and the specific function of each protein.

\subsection{Protein structures}
Proteins have four levels of structure:
\begin{itemize}
    \item \textbf{Primary Structure}: The sequence of amino acids in a polypeptide chain.
    \item \textbf{Secondary Structure}: Localized areas stabilized by hydrogen bonds, such as alpha-helices and beta-sheets.
    \item \textbf{Tertiary Structure}: The overall 3-dimensional structure of a single protein molecule; the spatial relationship of the secondary structures to one another.
    \item \textbf{Quaternary Structure}: The structure formed by several protein molecules (polypeptide chains), usually called protein subunits in this context, which function as a single protein complex.
\end{itemize}

\subsection{Homologous Proteins}
Homologous proteins are proteins that share a common evolutionary origin, indicated by their similarity in sequence or structure across different species. 
These proteins may retain similar functions, suggesting their conservation through evolutionary processes. 
Homologous proteins can be categorized into two types: orthologs, which are proteins in different species that evolved from a common ancestral gene and perform similar functions; 
and paralogs, which are proteins within the same species that arose from gene duplication and have evolved new functions. 
The study of homologous proteins helps in understanding protein function, evolution, and the genetic relationships between species.

\subsection{Mutually Exclusive Functions}
Mutually exclusive functions of proteins refer to situations where different protein isoforms, arising from the same gene through alternative splicing, perform distinct and non-overlapping roles within a cell or organism. 
Each isoform might be involved in different pathways or cellular processes, ensuring that their functions do not overlap, thus providing a mechanism for regulating biological activities and responses to environmental changes. 
This concept is crucial for understanding how genetic information can give rise to multiple functional outcomes, allowing a single gene to contribute to various aspects of cellular function and organismal development

\subsection{Dictionary of Protein Secondary Structure (DSSP)}
DSSP~\citep{DSSP} is a standard method used to describe the secondary structure of proteins, which has become one of the most widely used tools for analyzing protein structures.
DSSP analyzes the atomic coordinates of proteins to determine the secondary structure type of each amino acid residue, such as $\alpha$-helix, $\beta$-strand, turn, etc. It also provides additional information, such as hydrogen bonds between residues, the backbone torsion angles ($\phi$ and $\psi$), and solvent accessibility.

The output format of DSSP typically consists of a text file, with each line corresponding to an amino acid residue in the protein structure. Each line contains a series of labels describing the secondary structure and its local environment for that residue. Generally, the DSSP output format includes the following key information:

\begin{itemize}
\item \textbf{Residue Number}: Indicates the position number of each amino acid residue in the protein.
\item \textbf{Residue Type}: Specifies the type of amino acid residue, such as ALA, ARG, ASP, etc.
\item \textbf{Chain Identifier}: Indicates the identifier for different chains in the protein, usually a letter.
\item \textbf{Secondary Structure}: Indicates the secondary structure type for the residue, such as H ($\alpha$-helix), E ($\beta$-sheet), T (turn), etc.
\item \textbf{Hydrogen Bonds}: Describes the hydrogen bonding interactions between the residue and its neighboring amino acid residues.
\item \textbf{Phi ($\phi$) and Psi ($\psi$) angles}: Represents the backbone $\phi$ and $\psi$ angles for the residue, used to describe the protein's backbone conformation.
\item \textbf{Solvent Accessibility}: Describes the extent of exposure of the residue surface, i.e., the degree of contact between the residue and the aqueous solvent.
\end{itemize}

Each line in the DSSP output file contains a combination of these pieces of information to describe the features and environment of each residue in the protein structure.

\section{Vision-Language Pretraining (VLP)} \label{VLP}

Data collected from different modalities generally offer distinct perspectives, frequently synergizing to yield a comprehensive understanding, enhancing the overall comprehension of the data.
Vision-Language Pretraining (VLP) aims to learn multimodal foundation models, showing improved performance on various vision-and-language tasks~\citep{VLP}.
Existing VLP methods can be roughly divided into \textit{representation learning-based} and \textit{generative learning-based}.
\textit{Representation learning-based} methods~\citep{radford2021learning,jia2021scaling,yao2021filip,li2022grounded,li2021align} usually consider the image-text pairs as multi-modal views of the same semantics, and perform contrastive or multi-view learning for the alignment between multiple modalities. 
\textit{Generative learning-based} methods ~\citep{li2019visualbert,lu2019vilbert,chen2020uniter,li2020oscar,zhang2021vinvl,wang2022vlmixer,zeng2021multi,bao2022vl} aim to reconstruct the corrupted text (image) with the assistance of visual (text) modality through MLM-like objectives.
For example, SimVLM~\citep{wang2021simvlm} introduces a single prefix language modeling (PrefixLM) objective for exploiting large-scale weak supervision in VLP. 
CoCa~\citep{yu2022coca} further verifies the representation ability of autoregressive language modeling in the vision-language domain. 
In this paper, we consider protein as a specialized biological language that encodes and communicates biological information through its amino acid sequences and interactions.
Inspired by existing VLP methods~\citep{BLIP2,miniGPT4}, we first propose a representation learning-based Protein-Language Pretraining (PLP) framework to understand protein sequences via natural language instructions, and then 
design a generative learning-based multi-level adapter to generate the question-related answers combined with extra structural embeddings.

\section{Details on Protein-Language Representation Learning}
\label{PLPdetail}
Obtaining queries that can extract informative protein representation regarding text is significant for protein-language alignment. To achieve this, we connect our PLP-former with the ESM-1b~\citep{ESM-1b} model during the representation learning phase and train with the protein-language pairs. Following~\citep{BLIP,BLIP2}, we jointly train our model with three distinct pretraining tasks: Protein-Text Contrastive learning (PTC), Protein-grounded Text Generation (PTG), and Protein-Text Matching (PTM). Although these tasks utilize the same model structure and input format, they differ in the attention masking strategy applied between queries and text, thereby modulating their interaction. 

\paragraph{Protein-Text Contrastive Learning (PTC)}
For Protein-Text Contrastive Learning, by maximizing the mutual information, we aim to ensure the latent representation of protein and text are well-aligned. Specifically, given the query representation from the protein transformer $\mE_{seq}$, we align it with the corresponding text embedding $t$. This is achieved by maximizing the similarity of positive pairs against those negative pairs where we directly use the embedding of $[cls]$ tokens from the text transformer as $t$. Given that the output of the protein transformer comprises multiple embeddings, we calculate the pairwise similarity between each query output and $t$. We then choose the highest value to represent the protein-text similarity. To prevent any information leakage, we utilize an unimodal self-attention mask that restricts direct interaction between queries and text.

\paragraph{Protein-grounded Text Generation (PTG)}
The PTG task is designed to ensure that the learned queries can efficiently derive text-relevant information from the protein sequence. We train the PLP-former to produce descriptions matching the respective protein sequences. Since the PLP-former prevents direct interaction between the frozen ESM-1b and text tokens, the data needed for description generation must first be garnered by the queries, ensuring efficient information extraction. We use a multimodal causal self-attention mask, to manage the interaction between queries and text. While queries can interact with one another, they cannot engage with the text tokens. Conversely, each text token can reference all queries as well as its preceding text tokens. Additionally, we substitute the $[CLS]$ token with a $[DEC]$ token at the beginning of the text sequence to indicate the decoding task.

\paragraph{Protein-Text Matching (PTM)}
The protein-text Matching task is leveraged to align fine-grained protein-text representations. This task is designed as a binary classification task where the model needs to determine if a given image-text pair aligns (positive) or misaligns (negative). We employ a bi-directional self-attention mask, allowing all queries and texts to attend mutually. As a result, the obtained query embeddings, $\mE_{seq}$, encompass multimodal information. Each of these embeddings is then passed through a binary linear classifier to derive a logit, with the final matching score being the average of logits across all queries. For crafting informative negative pairs, we utilize the hard negative mining technique as described in~\citep{li2021align}.

\section{Details and Examples of the Training Datasets}
\label{appedix:dataset}
\textbf{ProtDescribe Dataset.}
The ProtDescribe dataset~\citep{protst} is designed to augment protein sequences with text descriptions of their functions and other important properties.
This dataset contains 549,000 pairs of proteins, including the EntryName, ProteinName, Function (obtained from Uniport), SubcellularLocation, Similarity, and ProteinSequence.
In our implementation, we only use the \textbf{protein sequence} and the corresponding \textbf{protein functions} for the training of PLP-former.
The PLP-former takes both sequence and descriptions as inputs, after three combined training tasks (Appendix.~\ref{PLPdetail}), then outputs 32 learned tokens indicating the sequence embeddings that are most relevant to the descriptions.
Some examples of this dataset are given in Table.~\ref{tab:protdescribe}.

\renewcommand\arraystretch{1.1}
\begin{table*}[t]
\caption{Examples of the original ProtDescribe dataset.
}\label{tab:protdescribe}
\scriptsize
\renewcommand\tabcolsep{1pt}
\centering
\begin{tabular}{l|l|l|l|l|l}
\hline
\multicolumn{1}{c}{EntryName} & \multicolumn{1}{c}{ProteinName} & \multicolumn{1}{c}{Function}  & \multicolumn{1}{c}{SubcellularLocation}    & \multicolumn{1}{c}{Similarity}   & \multicolumn{1}{c}{Sequence} \\ \hline
14KL\_BRUSI   & Lectin-like protein BA14k  & 
\begin{tabular}[c]{@{}l@{}}Has immunoglobulin-binding \\and hemagglutination  \\properties, and can bind to \\mannose. Essential for \\virulence. May be involved in\\ LPS biosynthesis or \\polysaccharide transport.\end{tabular} & \begin{tabular}[c]{@{}l@{}}Cell membrane; \\ Single-pass \\ membrane protein\end{tabular} & \begin{tabular}[c]{@{}l@{}}belongs to the \\BA14k family.\end{tabular}   & \begin{tabular}[c]{@{}l@{}}MNSFRKTCAGALA\\ LIFGATSIVPTVAAP\\ MNMDRPAINQNVI\\ QARAHYRPQNYNR\\ GHRPGYWHGHRG\\ YRHYRHGYRRHND\\ GWWYPLAAFGAGA\\ IIGGAISQPRPVYRAP\\ AG SPHVQWCYSRYK\\ SYRASDNTFQPYNGP\\ RKQCRSPYSR\end{tabular}   \\ \hline
11013\_ASFWA   & Protein MGF 110-13L   & \begin{tabular}[c]{@{}l@{}}Plays a role in virus cell \\ tropism, and may be required \\ for efficient virus replication\\ in macrophages.\end{tabular}    & \begin{tabular}[c]{@{}l@{}}Host membrane; \\ Multi-pass \\ membrane protein\end{tabular}  & \begin{tabular}[c]{@{}l@{}}Belongs to the \\asfivirus MGF \\110 family.\end{tabular}  & \begin{tabular}[c]{@{}l@{}}MGGGDYWPIIIRHCC\\ FYLVFSIAFVGYIVFA\\ YYKNLHLNTTMKLIA\\ LLCILIWLSQPGLNRP\\ LSIFYMKQNLPRTYTP\\ PIRELEYWCTYGKHC\\ DFCWECRNGICKNK\\ VWDDMPLIKQNDYIS\\ QCSIARYFDRCMYFIK\\ PKTPYIHYMDCSQPT\\ AYKGFSH\end{tabular} \\ \hline
\end{tabular}
\end{table*}

\renewcommand\arraystretch{1.1}
\begin{table*}[t]
\caption{Examples of the RCSB-PDB dataset with the protein entry ID from Protein Data Bank.
}\label{tab:RCSB}
\scriptsize
\renewcommand\tabcolsep{1pt}
\centering
\begin{tabular}{l|l}
\hline
\multicolumn{1}{c|}{Protein ID} & \multicolumn{1}{c}{Abstract of the Primary Publication} \\ \hline
1N21                            & \begin{tabular}[c]{@{}l@{}}The x-ray crystal structure of dimeric (+)-bornyl diphosphate synthase, a metal-requiring monoterpene cyclase from Salvia \\ officinalis, is reported at 2.0-A resolution. Each monomer contains two alpha-helical domains: the C-terminal domain \\catalyzes the cyclization of geranyl diphosphate, orienting and stabilizing multiple reactive carbocation intermediates; \\the N-terminal domain has no clearly defined function, although its N terminus caps the active site in the C-terminal domain \\during catalysis. Structures of complexes with aza analogs of substrate and carbocation intermediates, as well as complexes \\with pyrophosphate and bornyl diphosphate, provide "snapshots" of the terpene cyclization cascade.\end{tabular}  \\ \hline
2J11                            & \begin{tabular}[c]{@{}l@{}}The role of hydrophobic amino acids in the formation of hydrophobic cores as one of the major driving forces in protein \\folding has been extensively studied. However, the implication of neutral solvent-exposed amino acids is less clear and\\ available information is scarce. We have used a combinatorial approach to study the structural relevance of three solvent\\-exposed residues (Tyr(327), Thr(329), and Gln(331)) located in thebeta-sheet of the tetramerization domain of the tumor\\ suppressor p53 (p53TD). A conformationally defined peptide library was designed where these three positions were \\randomized. The library was screened for tetramer stability. A set of p53TD mutants containing putative stabilizing or\\ destabilizing residue combinations was synthesized for a thermodynamic characterization. Unfolding experiments showed \\a wide range of stabilities, with T(m) values between 27 and 83 degrees C. Wild type p53TD and some highly destabilized \\and stabilized mutants were further characterized. Thermodynamic and biophysical data indicated that these proteins were \\folded tetramers, with the same overall structure, in equilibrium with unfolded monomers. An NMR study confirmed that \\the main structural features of p53TD are conserved in all the mutants analyzed. The thermodynamic stability of the different \\p53TD mutants showed a strong correlation with parameters that favor the formation and stabilization of the beta-sheet. We \\propose that stabilization through hydrophobic interactions of key secondary structure elements might be the \\underlying mechanism for the strong influence of solvent-exposed residues in the stability of p53TD.\end{tabular} \\ \hline
3A08                            & \begin{tabular}[c]{@{}l@{}}Two crystal modifications of a collagen model peptide, (Pro-Pro-Gly)(4)-Hyp-Hyp-Gly-(Pro-Pro-Gly)(4) {[}where Hyp is \\(4R,2S)-L-hydroxyproline{]}, showed very similar unit-cell parameters and belonged to the same space group P2(1). Both \\crystals exhibited pseudo-merohedral twinning. The main difference was in their molecular-packing arrangements. One \\modification showed pseudo-hexagonal packing, while the other showed pseudo-tetragonal packing. Despite their \\different packing arrangements, no significant differences were observed in the hydration states of these modifications. \\The peptide in the pseudo-tetragonal crystal showed a cyclic fluctuation of helical twists with a period of 20 A, while that \\in the pseudo-hexagonal crystal did not. In these modifications, the puckering conformations of four of the 12 Hyp residues \\at the X position of the Hyp(X)-Hyp(Y)-Gly sequence were in the opposite conformations to the previous hypothesis that \\Hyp(X) residues involved in Hyp(X):Hyp(Y) and Hyp(X):Pro(Y) stacking pairs prefer up-puckering and down-puckering \\conformations, respectively. Detailed investigation of the molecular interactions between Hyp(X) and adjacent molecules \\revealed that these opposite conformations appeared because the puckering conformation, which follows the hypothesis, \\is subject to steric hindrance from the adjacent molecule.\end{tabular}   \\ \hline
\end{tabular}
\end{table*}

\renewcommand\arraystretch{1.1}
\begin{table*}[t]
\caption{An example of our instruction tuning dataset. We only show the input abstract and the generated instruction pairs for simplicity.
}\label{tab:bonito1}
\scriptsize
\centering
\begin{tabular}{l|l}
\hline
\multicolumn{1}{c|}{Abstract} & \multicolumn{1}{c}{Instruction pairs} \\ \hline
\begin{tabular}[c]{@{}l@{}} Assembly of polymerase chain reactions\\ at room temperature can sometimes lead\\ to low yields or unintentional products \\due to mispriming. Mutation of isoleucine \\707 to leucine in DNA polymerase I from\\ Thermus aquaticus substantially decreases\\ its activity at room temperature without \\compromising its ability to amplify DNA. \\To understand why a conservative change to\\ the enzyme over 20 u00c5 from the active \\site can have a large impact on its activity \\at low temperature, we solved the X-ray \\crystal structure of the large (5'-to-3' \\exonuclease-deleted) fragment of Taq DNA \\polymerase containing the cold-sensitive \\mutation in the ternary (E-DNA-ddNTP) and \\binary (E-DNA) complexes. The I707L \\KlenTaq1 ternary complex was identical to the \\wild-type in the closed conformation except \\for the mutation and a rotamer change in \\nearby phenylalanine 749, suggesting that the \\enzyme should remain active. However, soaking \\out of the nucleotide substrate at low temper-\\ature results in an altered binary complex made \\possible by the rotamer change at F749 near the \\tip of the polymerase O-helix. Surprisingly, two \\adenosines in the 5'-template overhang fill the \\vacated active site by stacking with the primer \\strand, thereby blocking the active site at low \\temperature. Replacement of the two overhanging \\adenosines with pyrimidines substantially \\increased activity at room temperature by keeping \\the template overhang out of the active site, \\confirming the importance of base stacking. These \\results explain the cold-sensitive phenotype of the \\I707L mutation in KlenTaq1 and serve as an example \\of a large conformational change affected by a \\conservative mutation. \end{tabular}
& \begin{tabular}[c]{@{}l@{}}Q: What is the effect of mutating isoleucine 707 to leucine in DNA polymerase \\I from Thermus aquaticus?\\ A: The mutation substantially decreases the enzyme's activity at room \\temperature without compromising its ability to amplify DNA.\\ Q: Why is it significant that the mutation site is over 20 Å away from \\the active site?\\ A: It is significant because it demonstrates that even conservative changes \\far from the active site can have a large impact on enzyme activity at low \\temperatures.\\ Q: What method was used to understand the impact of the I707L mutation on \\enzyme activity?\\ A: The X-ray crystal structure of the large fragment of Taq DNA polymerase \\containing the cold-sensitive mutation was solved in both ternary \\(E-DNA-ddNTP) and binary (E-DNA) complexes.\\ Q: How did the I707L mutation affect the structure of the ternary complex?\\ A: The I707L KlenTaq1 ternary complex was identical to the wild-type in the \\closed conformation, except for the mutation and a rotamer change in nearby \\phenylalanine 749.\\ Q: How does the I707L mutation affect the enzyme's activity at room \\temperature?\\ A: The I707L mutation substantially decreases the enzyme's activity at room \\temperature.\\ Q: What happens to the I707L mutation's nucleotide substrate at low \\temperatures?\\ A: The nucleotide substrate soaks out at low temperatures, resulting in an \\altered binary complex.\\ Q: What specific mutation was made to DNA polymerase I?\\ A: Isoleucine 707 was mutated to leucine in DNA polymerase I.\\ Q: What structural method was used to analyze the I707L mutation's impact?\\ A: X-ray crystallography was used to determine the structure of the mutated \\enzyme.\\ Q: What was observed in the ternary complex structure of the I707L mutated \\enzyme?\\ A: The ternary complex structure was identical to the wild-type except for the \\mutation and a rotamer change in phenylalanine 749.\\ Q: What effect does the rotamer change at phenylalanine 749 have on the \\enzyme?\\ A: The rotamer change enables two adenosines from the 5'-template overhang \\to fill the vacated active site, blocking it at low temperatures.\end{tabular} \\ \hline
\end{tabular}
\end{table*}

\renewcommand\arraystretch{1.1}
\begin{table*}[t]
\caption{Another example of our instruction tuning dataset.
}\label{tab:bonito2}
\scriptsize
\centering
\begin{tabular}{l|l}
\hline
\multicolumn{1}{c|}{Abstract} & \multicolumn{1}{c}{Instruction pairs} \\ \hline
\begin{tabular}[c]{@{}l@{}} The bacterial ubiD and ubiX or the\\ homologous fungal fdc1 and pad1 \\genes have been implicated in the \\non-oxidative reversible decarboxy-\\lation of aromatic substrates, and \\play a pivotal role in bacterial ubi-\\quinone (also known as coenzyme \\Q) biosynthesis or microbial bio-\\degradation of aromatic compounds, \\respectively. Despite biochemical \\studies on individual gene products, \\the composition and cofactor require-\\ment of the enzyme responsible for \\in vivo decarboxylase activity \\remained unclear. Here we show that \\Fdc1 is solely responsible for the \\reversible decarboxylase activity, and\\ that it requires a new type of cofactor: \\a prenylated flavin synthesized by the \\associated UbiX/Pad1. Atomic resolu-\\tion crystal structures reveal that two \\distinct isomers of the oxidized co-\\factor can be observed, an isoalloxazine \\N5-iminium adduct and a N5 secondary \\ketimine species with markedly altered \\ring structure, both having azomethine \\ylide character. Substrate binding posi-\\tions the dipolarophile enoic acid group\\ directly above the azomethine ylide \\group. The structure of a covalent \\inhibitor-cofactor adduct suggests that \\1,3-dipolar cycloaddition chemistry \\supports reversible decarboxylation in \\these enzymes. Although 1,3-dipolar \\cycloaddition is commonly used in \\organic chemistry, we propose that this \\presents the first example, to our know-\\ledge, of an enzymatic 1,3-dipolar \\cycloaddition reaction. Our model for \\Fdc1/UbiD catalysis offers new routes \\in alkene hydrocarbon production or \\aryl (de)carboxylation. \end{tabular}
& \begin{tabular}[c]{@{}l@{}}Q: What enzymatic activity is Fdc1 responsible for?\\ A: Fdc1 is responsible for the reversible decarboxylation of aromatic substrates.\\ Q: What type of cofactor does Fdc1 require for its activity?\\ A: Fdc1 requires a prenylated flavin cofactor for its activity.\\ Q: How is the cofactor for Fdc1 synthesized?\\ A: The cofactor for Fdc1 is synthesized by UbiX/Pad1.\\ Q: What are the two isomeric forms of the oxidized cofactor in Fdc1?\\ A: The two isomeric forms are the isoalloxazine N5-iminium adduct and the N5 \\secondary ketimine species.\\ Q:What characterizes the cofactor forms in Fdc1?\\ A: Both cofactor forms in Fdc1 have azomethine ylide character.\\ Q: How does substrate binding influence Fdc1's catalytic mechanism?\\ A: Substrate binding positions the enoic acid group directly above the azomethine \\ylide group, facilitating the catalytic process.\\ Q: What does the covalent inhibitor-cofactor adduct structure in Fdc1 suggest about \\its mechanism?\\ A: It suggests that 1,3-dipolar cycloaddition chemistry is a mechanism supporting\\ reversible decarboxylation in Fdc1.\\ Q: What makes Fdc1's reaction significant in enzymatic chemistry?\\ A: Fdc1's reaction may represent the first enzymatic example of 1,3-dipolar cycloaddition, \\according to the text.\\ Q: What potential industrial application does Fdc1's enzymatic activity suggest?\\ A: Fdc1's enzymatic activity suggests potential applications in alkene hydrocarbon \\production and aryl (de)carboxylation processes.\\ Q: What does the altered ring structure of the cofactor in Fdc1 indicate about its function?\\ A: The altered ring structure indicates significant changes in electronic structure that \\likely influence the enzyme's catalytic activity.\end{tabular} \\ \hline
\end{tabular}
\end{table*}

\textbf{RCSB-PDB Dataset.}
The RCSB-PDB dataset~\citep{ProteinChat} is originally sourced from the Research Collaboratory for Structural Bioinformatics Protein Data Bank~\footnote{\url{https://www.rcsb.org}}, which includes 204,826 experimentally
determined 3D structures.
ProteinChat~\citep{ProteinChat} further selected 143,508 proteins that have a primary publication linked with a PubMed ID and can be extracted with a valid chain.
Some examples of this dataset are given in Table.~\ref{tab:RCSB}.

\textbf{Instruction Tuning Dataset.}
Our refined instruction tuning dataset is based on the RCSB-PDB dataset. The RCSB-PDB dataset contains a protein entry ID from Protein Data Bank, a pdb file with atomic coordinates representing the molecular
structure, along with the abstract of corresponding scientific literature.
In this paper, we extract the amino acid chain as the primary sequence and further use the abstract to generate 10 Q\&A pairs for instruction tuning.
In summary, for every protein taken into account, we compile its residue sequences, secondary structure label, tertiary atomic coordinates, along with corresponding 10 paired Q\&A descriptions.
We use the pretrained Bonito~\citep{Bonito}, an open-source model~\footnote{\url{https://huggingface.co/BatsResearch/bonito-v1}} for conditional task generation from specialized domains across three task types: yes-no question answering, extractive question answering, and natural language inference.
Bonito takes the unannotated text and the task attribute as inputs, and the output consists of the instruction and the response. 
In this paper, we use Bonito to generate synthetic Q\&A pairs based on the corresponding scientific literature (only abstract used for simplicity). In our implementation, the task attribute is set as `` extractive question answering''.
Some examples of our refined dataset are given in Table.~\ref{tab:bonito1} and Table.~\ref{tab:bonito2}.

\section{More Implementation Details}\label{implementation}
During training, the PLP-former is initialized with the pre-trained weights of PubMedBERT~\cite{pubmedbert}, whereas the cross-attention layers and learnable tokens are randomly initialized.
Following the same setting as PubMedBERT, we use 32 learnable tokens with a dimension of 768, which is the same as the hidden dimension of the PLP-former. 
We pre-train the PLP-former on the ProtDescribe dataset for 20K epochs with a batch size of 64.
We use the AdamW~\citep{adamw} optimizer with $\beta_1$ = 0.9, $\beta_2$ = 0.98, and a weight decay of 0.05. 
We use a cosine learning rate decay with a peak learning rate of 1e-4, a minimum learning rate of 8e-5, and a linear warm-up of 5K iterations. 
Note that we apply PLP only to the sequence embeddings since the reported protein structures are much less than sequences. 
For example, there are 182K experimentally determined structures in PDB~\citep{PDB} while 47M protein sequences in Pfam~\citep{pfam}.
Thus we only use selected sequence-structure pairs from the RCSB-PDB dataset during the following stage.
During contrastive learning of our Protein Context Gating (PCG) module, the temperature parameter $\tau$ is empirically set to 0.8, and $k$ is set to 128 as the number of both positive and negative samples.
Finally, we freeze the PLP-former, PCG module and LLM decoder, and train the projection adapter for 1K epochs with a batch size of 128. 
The minimum learning rate for the second stage is set as 5e-5.
We also conduct experiments on fine-tuning LLMs, evaluating both full parameter tuning and LoRA~\citep{lora} tuning on Vicuna-13B. Additionally, we fine-tuned LL{\scalebox{0.75}[0.75]{A}}MA3 with LoRA for comparison.
To avoid out-of-memory issues, we randomly trim excessively long protein sequences and structures to 3,000 amino acids.
All experiments are performed on 2 NVIDIA H-100 (80GB) GPUs.
Our model with Vicuna-13B requires 3 days for the multi-level protein-language alignment stage and 2.5 days for the instruction tuning stage (full-parameter fine-tuning), while the LL{\scalebox{0.75}[0.75]{A}}MA3-8B model takes approximately 1 day for the instruction tuning stage (LoRA fine-tuning).
The pre-trained Vicuna-13B and LL{\scalebox{0.75}[0.75]{A}}MA3-8B are obtained through the official repositories\footnote{\url{https://github.com/meta-llama/llama3}}\textsuperscript{,}\footnote{\url{https://huggingface.co/lmsys/vicuna-7b-v1.5}}.

\section{Additional Experiments}\label{moreex}
\subsection{More results on LLM fine-tuning}
In our implementation, we use the frozen LLM encoder as default since the projection adapter works as an information bottleneck between hybrid protein features and the LLM decoder.
We further improve the performance by deploying instruction tuning on the LLMs, as described in Section~\ref{quan}.
We perform 16-bit LoRA~\citep{lora} on Vicuna-13B and quantized 4-bit QLoRA~\citep{qlora} on LL{\scalebox{0.75}[0.75]{A}}MA3-8B. We also perform full-parameter funding on the Vicuna-13B model.
Detailed results are shown in Table.~\ref{tab:appendix}.
\renewcommand\arraystretch{1.1}
\begin{table*}[t] 
    \caption{Quantitative Comparisons on fine-tuned LLM decoders. $\uparrow$ indicates that a higher value corresponds to better performance. FP means full parameter tuning.
    }\label{tab:appendix}
    \scriptsize
    \renewcommand\tabcolsep{6pt}
    \centering
     \vspace{-2mm}
    \begin{tabular}{l|ccccccc}
    \hline
    \textbf{Method}     & \textbf{BLEU-1 $\uparrow$} & \textbf{BLEU-4 $\uparrow$} & \textbf{ROUGE-L $\uparrow$} & \textbf{METEOR $\uparrow$} & \textbf{CIDEr $\uparrow$} & \textbf{SPICE $\uparrow$} & \begin{tabular}[c]{@{}c@{}}\textbf{PubMed}\\ \textbf{BERTScore $\uparrow$}\end{tabular} \\ \hline
    ProtChatGPT (Vicuna-13B)    & \textbf{0.618}  & \textbf{0.399}  & \textbf{0.493}   & \textbf{0.299}  & \textbf{0.640} & \textbf{0.321} & \textbf{0.465}\\ 
    ProtChatGPT (LL{\scalebox{0.75}[0.75]{A}}MA3-8B)    & \textbf{0.643}  & \textbf{0.417}  & \textbf{0.515}   & \textbf{0.326}  & \textbf{0.687} & \textbf{0.346} & \textbf{0.487}
    \\
    ProtChatGPT (Vicuna-13B-LoRA)    & \textbf{0.670}  & \textbf{0.435}  & \textbf{0.547}   & \textbf{0.344}  & \textbf{0.695} & \textbf{0.377} & \textbf{0.505}  \\ 
    ProtChatGPT (LL{\scalebox{0.75}[0.75]{A}}MA3-8B-QLoRA)    & \textbf{0.698}  & \textbf{0.453}  & \textbf{0.556}   & \textbf{0.354}  & \textbf{0.746} & \textbf{0.391} & \textbf{0.529}  
    \\
    ProtChatGPT (Vicuna-13B-FP)    & \textbf{0.735}  & \textbf{0.479}  & \textbf{0.603}   & \textbf{0.379}  & \textbf{0.761} & \textbf{0.416} & \textbf{0.557}
    \\ \hline
    \end{tabular}
\end{table*}

\subsection{Experiments on Protein Retrieval}
\begin{figure*}[t]
	\centering
	\includegraphics[width=0.9\textwidth]{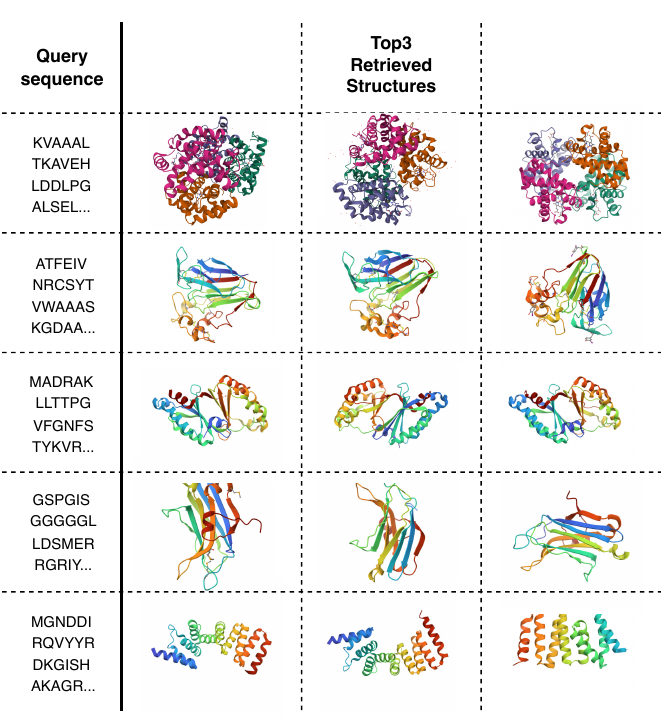}
	\vspace{-2mm}
	\caption{Results on cross-level protein retrieval task.}
        \label{fig:retrieval}
\end{figure*}

In this paper, we design a Protein Context Gating (PCG) module to align the protein sequence with secondary structure embedding, and further use contrastive learning to align the tertiary structures.
To validate the effectiveness of our multi-level protein alignment, we conducted cross-level protein retrieval experiments. Specifically, we randomly selected sequences from the test set, processed them through the PCG module to obtain $\mE_{align}$, and then identified the closest protein by calculating the cosine similarity between $\mE_{align}$ and the tertiary structure embedding $\mE_{ter}$ of other proteins. The results are shown in Figure~\ref{fig:retrieval}.

\section{Details on Metrics}
\label{metrics}

In our implementation, we use seven different metrics on the proposed ProtChatGPT to verify the performance. These metrics focus not only on low-level lexical and syntactic alignment but also on high-level semantic information. 

\textbf{BLEU}~\citep{papineni2002bleu} (BiLingual Evaluation Understudy) serves as a crucial metric for assessing the quality of the machine-generated text, particularly in machine translation contexts. It quantifies the similarity between the candidate and reference text, yielding a score within the range of 0 to 1. A higher BLEU score indicates a closer match between the candidate and reference texts.
BLEU is mathematically defined as follows:
\begin{equation}
    p_n = \frac{\sum\limits_{C \in \{Candidates\}} \sum\limits_{\textit{n-gram} \in C} Count_{clip} (\textit{n-gram})}{\sum\limits_{C^{\prime} \in \{Candidates\}} \sum\limits_{\textit{n-gram}^{\prime} \in C^{\prime}} Count (\textit{n-gram}^{\prime})},
\end{equation}
\begin{equation}
    \text{BP} = 
    \begin{cases}
        1 & \text{if } c > r \\
        e^{(1-r/c)} & \text{if } c \le r,
    \end{cases}
\end{equation}
\begin{equation}
    \text{BLEU} = \text{BP} \cdot exp \left(\sum_{n=1}^{N} w_n log p_n \right),
\end{equation}
where BP (Brevity Penalty) helps penalize overly short translations and $p_n$ represents the precision of n-grams, ranging from 1 to a predefined maximum.

\textbf{ROUGE-L}~\citep{lin2002manual} finds frequent use in the automatic evaluation of text summarization and machine translation. It calculates the longest common subsequence between the candidate and reference texts, prioritizing recall over precision.
ROUGE-L can be formulated as:
\begin{equation}
    R_{lcs} = \frac{LCS(X, Y)}{m},
\end{equation}
\begin{equation}
    P_{lcs} = \frac{LCS(X, Y)}{n},
\end{equation}
\begin{equation}
    \text{ROUGE-L} = \frac{(1+\beta^2)R_{lcs}P_{lcs}}{R_{lcs}+\beta^2P_{lcs}},
\end{equation}
where $X$ represents the predicted text with a length of $n$. $Y$ represents the ground truth text with a length of $m$. $\beta$ is a hyperparameter used to adjust the emphasis on precision and recall.
LCS calculates the length of the longest common subsequence, $R_{lcs}$ measures recall, and $P_{lcs}$ measures precision, respectively.

\textbf{METEOR}~\citep{banerjee2005meteor} provides a comprehensive evaluation of machine-generated text by considering not only exact word matches but also synonyms and stemming. It combines precision, recall, and alignment factors to offer a holistic assessment.
METEOR is formulated as:
\begin{equation}
    F=\frac{(\alpha^2+1)P}{R+\alpha P},
\end{equation}
\begin{equation}
    \text{Meteor}=(1-Penalty)\cdot F,
\end{equation}
where $Penalty$ penalizes excessive word mismatches, $\alpha$ is a configurable parameter, $R$ and $P$ represents recall and precision respectively.

\textbf{CIDEr}~\citep{vedantam2015cider} (Consensus-based Image Description Evaluation) primarily assesses the quality of image captions produced by automated systems. It places importance on consensus among multiple reference captions and emphasizes the inclusion of diverse descriptive words.
The mathematical formulation of CIDEr is given by:
\begin{equation}
    CIDEr_n(c,S)=\frac1M\sum_{i=1}^M\frac{g^n(c)\cdot g^n(S_i)}{||g^n(c)||\times||g^n(S_i)||},
\end{equation}
where $c$ represents the candidate text, $S$ denotes the set of reference texts, $n$ specifies the use of n-grams, $M$ represents the number of reference texts, and $g$ corresponds to the TF-IDF vector based on n-grams.

\textbf{SPICE}~\citep{spice2016} offers a metric designed to evaluate the semantic content of image captions, with a focus on their precision within generated captions.
The formulation of SPICE is as follows:
\begin{equation}
    P(c,S) = \frac{|T(G(c))\otimes T(G(S))|}{|T(G(c))|},
\end{equation}
\begin{equation}
    R(c,S) = \frac{|T(G(c))\otimes T(G(S))|}{|T(G(S))|},
\end{equation}
\begin{equation}
    SPICE(c,S) = \frac{2\cdot P(c,S)\cdot R(c,S)}{P(c,S)+R(c,S)},
\end{equation}
where the binary matching operator $\otimes$ is the function that returns matching tuples in two scene graphs, $P$ represents the precision of semantic propositions and $R$ signifies the recall of semantic propositions, respectively.

\textbf{BertScore}~\citep{zhang2019bertscore} is a metric that leverages contextual embeddings from BERT models to assess the quality of machine-generated text. It measures the similarity between the candidate text and the reference text using contextual embeddings.

\textbf{PubMed BERTScore} comes from the classical \textbf{BertScore}~\citep{zhang2019bertscore}, which is a metric that leverages contextual embeddings from BERT models to assess the quality of machine-generated text. It measures the similarity between the candidate text and the reference text using contextual embeddings.
In our implementation, to better assess the quality of ProtChatGPT in generating protein-related descriptions, we further replaced the encoder with the encoder of PubMedBERT~\citep{pubmedbert}.
PubMedBERT is the latest BERT~\citep{bert} model pre-trained on the biomedical corpus, which outperformed BioBERT on the BLURB~\citep{pubmedbert} (Biomedical Language Understanding and Reasoning Benchmark).









\end{document}